\documentclass[%
 aip,
 amsmath,amssymb,
 reprint,%
]{revtex4-1}

\usepackage{graphicx}
\usepackage{dcolumn}
\usepackage{bm}

\usepackage[a4paper, left = 1.22cm, right = 1.22cm, top = 1.8cm, bottom = 1.8cm]{geometry}
\usepackage[utf8]{inputenc}
\usepackage[T1]{fontenc}
\usepackage{etoolbox}
\usepackage{color}
\usepackage{soul}
\usepackage{array,multirow}
\usepackage{xr-hyper}
\usepackage{hyperref}
\hypersetup{hidelinks}

\makeatletter
\def\@email#1#2{%
 \endgroup
 \patchcmd{\titleblock@produce}
  {\frontmatter@RRAPformat}
  {\frontmatter@RRAPformat{\produce@RRAP{*#1\href{mailto:#2}{#2}}}\frontmatter@RRAPformat}
  {}{}
}%
\makeatother

\begin{document}

\preprint{AIP/123-QED}

\title[Nudged Elastic Band Method in the \textsc{Crystal} code. Theory and Applications]{Nudged Elastic Band Method in the C{\normalsize RYSTAL} code. Theory and Applications}

\author{Andreha Gelli$^*$}
\email{andreha.gelli@unito.it, chiara.ribaldone@unito.it}
\affiliation{\footnotesize Dipartimento di Chimica, University of Torino, via Giuria 5, 10125, Torino, Italy}
\affiliation{\footnotesize Department de Química, Universitat Autònoma de Barcelona, E-08193, Catalonia, Spain}

\author{Silvia Casassa}

\affiliation{\footnotesize Dipartimento di Chimica, University of Torino, via Giuria 5, 10125, Torino, Italy}

\author{Albert Rimola}%
\affiliation{\footnotesize Department de Química, Universitat Autònoma de Barcelona, E-08193, Catalonia, Spain}
\affiliation{\footnotesize Accademia delle Scienze di Torino, Via Maria Vittoria, 3, 10123 Torino, Italy}

\author{Chiara Ribaldone$^*$}%
\affiliation{\footnotesize Dipartimento di Chimica, University of Torino, via Giuria 5, 10125, Torino, Italy}


\begin{abstract}

\noindent
The nudged elastic band ({\small NEB}) method is a widely used algorithm for determining minimum energy paths and transition states in chemical reactions and phase transitions. In this work, we present the implementation of different {\small NEB} algorithm schemes in the \textsc{Crystal} code, a quantum mechanical ab initio program for the calculation of electronic properties of condensed matter systems, based on Hartree-Fock and density functional theory. The use of a set of localized Gaussian-type functions to expand the wavefunction permits a very efficient evaluation of the exact exchange series for the hybrid exchange-correlation functionals. Therefore, our implementation allows an accurate characterization of transition states in both molecular and condensed phase systems, at the hybrid functional level of theory. The theoretical framework, including the force projection scheme, tangent estimation, optimization strategies, as well as the extensions of the method with climbing image and variable spring constant variants, is recalled. Then, the {\small NEB} algorithm is validated through a series of benchmark tests: two molecular reactions (a collinear proton transfer process and the keto-enol tautomerization in formamide) and a proton exchange process in a periodic chabazite zeolite. Our results are in excellent agreement with experimental and previous theoretical data, confirming the accuracy and applicability of the implementation. This work opens the possibility for future studies of complex reactive processes on extended periodic systems, using hybrid functionals.
\end{abstract}

\maketitle

%


\section{Introduction}

The calculation of energy barriers and transition states for processes such as surface-catalyzed heterogeneous reactions, ion trapping and migration within solid matrices, defect formation and propagation, as well as phase transitions, is a current challenge in condensed matter physics and theoretical chemistry. Despite the great advances in the computational community for the development of novel methods and algorithms to accelerate molecular dynamics simulations,\cite{herbert_md, planewave_md, lipparini_md_1, lipparini_md_2} such phenomena are typically many orders of magnitude slower than vibrations of the atoms, making direct ab initio dynamical simulations an impractical way to calculate transition rates. At the same time, this separation of time scales allows the use of statistical approaches such as the harmonic approximation to transition state theory ({\small HTST})\cite{wigner_tst, vineyard_htst} or Kramers theory.\cite{kramers_tst} Within these methods, the activation energy of a transition, representing the main factor that determines thermal stability, is defined as the energy difference between the local minimum of the potential energy surface corresponding to the initial state and the first-order saddle point along the path connecting the minima of the initial and final states.
Among all the families of theoretical methods and algorithms that have been developed for tackling the problem of saddle points searching on potential energy surfaces, the nudged elastic band ({\small NEB}) approach, introduced by Jónsson and Henkelman,\cite{jonsson_book, henkelman_1999, henkelman_new_tangent, henkelman_CIneb, sheppard_opt, sheppard_solidstate} has been established, since its early days, as one of the most efficient and accurate. Both its basic\cite{jonsson_book} and improved versions\cite{henkelman_new_tangent, henkelman_CIneb} are based on interpolation between the initial and final states of transition, converting the search for a first-order saddle point in configuration space to an optimization problem in a discretized path space. In this way, only the evaluation of the first order energy derivative is required, differently from surface walking approaches\cite{cerjan_walking, banerjee_walking, baker_walking} or transition states localization techniques.\cite{chab_rimola_2010}
Thanks to its efficiency, the {\small NEB} method is widely used in conjunction with electronic structure calculations such as Hartree-Fock or density functional theory ({\small DFT}), which allows for an accurate evaluation of the forces among atoms, as well as a precise estimation of the potential energy surface ({\small PES}). 
In particular, in the framework of {\small DFT}, semilocal density functionals can often be insufficient to achieve a good description of electronic properties associated with the {\small PES}, especially for localized defects or systems with $d$ and $f$ orbitals. The over-delocalization of the electrons, due to the intrinsic self-interaction error in the semilocal functionals, as well as the associated underestimation of the band gap, can be mitigated by introducing hybrid density functionals, which incorporate a fraction of nonlocal exact exchange.
Unfortunately, when studying extended systems, the evaluation of the exact exchange terms can become very demanding in terms of computational time, especially for plane waves-based codes.
In the \textsc{Crystal} code, the use of localized Gaussian functions as basis set for expanding the wavefunction has allowed the development of very efficient algorithms for the evaluation of the exact exchange series.\cite{crystal17, crystal23} 
Therefore, the current implementation, combining the accuracy of the saddle-point searching technique via {\small NEB} with the efficiency of the program in evaluating the exchange series, can pave the way for a high-accurate description of transitions states at the {\small DFT} level, by means of hybrid density functionals. 

The article is organized as follows.
In Section \ref{sec:theory}, the main theoretical background underlying the nudged elastic band method is briefly recalled and discussed. In Section \ref{sec:results}, with reference to selected test cases, ($i$) the reliability of the nudged elastic band method is assessed through comparisons with experimental data and ($ii$) its efficiency is analyzed in relation to other ab initio quantum mechanical programs. Finally, in Section \ref{sec:conclusions}, the conclusions are outlined, together with some future perspectives.

\section{Theory}\label{sec:theory}


Although the theoretical framework and methodological aspects are well known and consolidated, a dedicated section to introduce the current implementation of the {\small NEB} method is presented, so that users can get the most out of it, having documented the meaning and effect of the few parameters to be set and the specific of the various options.
For clarity, a flowchart of the algorithm is provided in Section 6
of the supplementary material. \\

Consider a given nuclear configuration as a collection of $N$ atoms in a $d$ dimensional space, described by a $Nd$ dimensional vector $\bm{q}_i \in \mathbb{R}^{Nd}$ where each component of the vector identifies a position of an atom in a given spatial direction. 
Using the concept of {\small PES}, a map can be established between each nuclear configuration $\bm{q}_i$ and its ground state energy $E(\bm{q}_i)$, computed using Hartree-Fock or {\small DFT}. \\
Starting from an initial state $\bm{q}_1$, the {\small NEB} method defines a scheme to find the minimum energy path connecting the initial state $\bm{q}_1$ with a specific final state $\bm{q}_m$, by sampling the {\small PES} with a set of $(m - 2)$ intermediate nuclear configurations. The total set of $m$ configurations thus defined, called images, can be indicated as $\bm{q} = [\bm{q}_1, \bm{q}_2,...,\bm{q}_m]$, where the endpoints $\bm{q}_1$ and $\bm{q}_m$ are fixed and correspond to two local minima of the {\small PES}.
The positions of the $(m - 2)$ intermediate images, which ultimately identify the minimum energy path ({\small MEP}), are found by minimizing the objective function
\begin{equation}\label{eq1}
    S(\bm{q}) = \sum_{i=1}^{m} E(\bm{q}_i) + \frac{1}{2}\sum_{i=2}^{m} k_{i(i-1)} \, (\bm{q}_i - \bm{q}_{i-1})^2
\end{equation}
This function represents a physical system consisting of an elastic chain formed by $m$ objects,
each connected to its neighbor by a spring with elastic constant $k_{i(i-1)}$, with the endpoints $\bm{q}_1$ and $\bm{q}_m$ held fixed. Therefore, the first term in equation \eqref{eq1} is the sum of the ground state energies of each nuclear configuration, while the second term is the elastic potential energy due to the presence of fictitious springs connecting the images.

\subsection{Generation of the initial path}\label{sec:initial_path}

Firstly, the question arises of how to define the initial guess for the intermediate nuclear configurations [$\bm{q}_2,...,\bm{q}_{m-1}$] which enter in the objective function \eqref{eq1} and represent the initial path between the endpoints in the {\small PES}. A simple yet effective solution is to initialize the intermediate images by means of a linear interpolation between the initial $\bm{q}_1$ and final $\bm{q}_m$ nuclear configurations. For completeness, the adopted algorithm is detailed in Section 1
of the supplementary material.

\subsection{Forces acting on the intermediate images}

Once the initial images are set, the minimization of the objective function \eqref{eq1}, that is, the search for its stationary points, provides the equation of motion for the displacement of each image towards the {\small MEP}.
Taking the derivative of equation \eqref{eq1} with respect to a given nuclear configuration, the form of the force acting on each nuclei in that image can be devised as
\begin{equation}\label{eq2}
    \bar{\mathbf{F}}_i^{\text{\tiny NEB}} = - \frac{\partial S(\bm{q})}{\partial \bm{q}_i} = \mathbf{F}_i + \mathbf{F}_i^s
\end{equation}
Following this prescription, the nuclear forces \eqref{eq2} will include two terms, one related to the real nuclear forces $\mathbf{F}_i = - \nabla E(\bm{q}_i)$, and the other one associated to the elastic forces $\mathbf{F}_i^s$ due to the spring interactions. Hence, the minimization of the total forces \eqref{eq2} acting on each image leads the initial chain of images to converge to the {\small MEP}, whose first-order saddle points are assigned with transition states, corresponding to the highest energy points on the {\small MEP} connecting the initial $\bm{q}_1$ and final $\bm{q}_m$ nuclear configurations.
However, this formulation implies two well-known problems: ($i$) the component of the real forces parallel to the path causes the images to move away from high energy regions towards the energy minima, thereby decreasing the density of the images in the neighborhood of the saddle points; ($ii$) the component of the elastic forces perpendicular to the path tends to prevent the band from following a curved trajectory, thus avoiding the retrieval of the true saddle point. 
The first issue, called the \emph{sliding down} problem, becomes relevant for small values of the elastic constant, and can be avoided considering only the component $\left.(\mathbf{F}_i)\right\vert_\perp$ of the real forces perpendicular to the path. The second challenge, known as \emph{corner cutting} effect, mostly arises for high values of the elastic constant, and can be overcome by taking into account only the component $(\mathbf{F}_i^s)|_\parallel$ of the elastic forces parallel to the path, thus leading to the following redefinition of the total forces acting on the $i$-th image
\begin{equation}\label{eq3}
\begin{aligned}
    \mathbf{F}_i^{\text{\tiny NEB}} & = \left.(\mathbf{F}_i)\right\vert_\perp \,\, + \,\, (\mathbf{F}_i^s)|_\parallel \\
    & = - \left.\nabla E(\bm{q}_i)\right\vert_\perp \,\, + \,\, (\mathbf{F}_i^s)|_\parallel 
\end{aligned} \quad \quad i = 2,...,(m-1) \,\,
\end{equation}

In order to evaluate the components of the forces with respect to the path, a local tangent unit vector $\hat{\boldsymbol{\tau}}_i$ has to be defined for each $i$-th image, so that the real and elastic forces can be projected according to equation \eqref{eq3}. In this way, the component of the real forces perpendicular to the path can be obtained by subtracting the component parallel to the path to the total real forces vector, that is
\begin{equation}\label{eq4}
\begin{aligned}
    \left.(\mathbf{F}_i)\right\vert_\perp \, & = \mathbf{F}_i - [\mathbf{F}_i \cdot \hat{\boldsymbol{\tau}}_i] \, \hat{\boldsymbol{\tau}}_i \\
    & = - \nabla E(\bm{q}_i) + [\nabla E(\bm{q}_i)\cdot \hat{\boldsymbol{\tau}}_i] \, \hat{\boldsymbol{\tau}}_i
\end{aligned} 
\end{equation}
while the component of the elastic forces parallel to the path can be computed using the well-known formula\cite{henkelman_new_tangent}
\begin{equation}\label{eq5}
    (\mathbf{F}_i^s)|_\parallel \, = \, k(\lVert\bm{q}_{i+1} - \bm{q}_i\rVert - \lVert\bm{q}_{i} - \bm{q}_{i-1}\rVert) \, \hat{\boldsymbol{\tau}}_i
\end{equation}
where $k = k_{(i+1)i} = k_{i(i-1)}$ is the elastic constant, set to an equal value $k$ for all the interactions among images.
Applying equations \eqref{eq4} and \eqref{eq5} for $i = 2,...,(m-1)$, that is, for all the intermediate images, and then using the formula \eqref{eq3}, the total nudged elastic band forces can be evaluated and minimized, thus finding the {\small MEP}.

\subsection{Estimation of the tangent to the path}

Initially, the calculation of $\hat{\boldsymbol{\tau}}_i$ was based on a simple estimation of the unit tangent at the $i$-th image as the normalized line segment between two configurations adjacent the $i$-th image.\cite{jonsson_book} Later, the evaluation of this quantity was made more effective, leading to better localization of the saddle points and avoiding the formation of kinks in the path, which cause instabilities and slow down convergence behavior, by defining it in relation to the energy of the first closest images, in the following way\cite{henkelman_new_tangent}
\begin{align}\label{eq6}
    & \boldsymbol{\tau}_i =
	\bigg \{
	\begin{array}{rl}
	\boldsymbol{\tau}_i^+ & \,\, \text{ if } \,\, E_{i+1} > E_i > E_{i-1} \\
	\boldsymbol{\tau}_i^- & \,\, \text{ if } \,\, E_{i+1} < E_i < E_{i-1} \\
	\end{array} \qquad i = 2,...,(m-1) \notag \\
	& \text{with} \quad \boldsymbol{\tau}_i^+ = \bm{q}_{i+1}-\bm{q}_{i} \quad \text{ and } \quad \boldsymbol{\tau}_i^- = \bm{q}_{i}-\bm{q}_{i-1} 
\end{align}
where $E_i \equiv E(\bm{q}_{i})$ is the ground state energy of the $i$-th image.
The previous equation \eqref{eq6} describes the case where adjacent images have monotonically increasing or decreasing energies. 
For cases where the $i$-th image is at a minimum ($E_{i+1} > E_i < E_{i-1}$) or at a maximum ($E_{i+1} < E_i > E_{i-1}$), then the tangent is estimated to be a linear combination of the vectors defined in equation \eqref{eq6}, leading to the expression\cite{henkelman_new_tangent}
\begin{equation}\label{eq7}
    \boldsymbol{\tau}_i =
	\bigg \{
	\begin{array}{rl}
	\boldsymbol{\tau}_i^+\Delta E_i^{\,\text{max}} + \boldsymbol{\tau}_i^-\Delta E_i^{\,\text{min}} & \,\, \text{ if } \,\, E_{i+1} > E_{i-1} \\
	\boldsymbol{\tau}_i^+\Delta E_i^{\,\text{min}} + \boldsymbol{\tau}_i^-\Delta E_i^{\,\text{max}} & \,\, \text{ if } \,\, E_{i+1} < E_{i-1} \\
	\end{array} 
\end{equation}
where the weight coefficients are given by
\begin{equation}\label{eq8}
	\begin{array}{rl}
	\Delta E_i^{\,\text{max}} &= \text{max}(\left|E_{i+1} - E_i\right|, \left|E_{i-1} - E_i\right|) \\
	\Delta E_i^{\,\text{min}} &= \text{min}(\left|E_{i+1} - E_i\right|, \left|E_{i-1} - E_i\right|) \\
	\end{array}
\end{equation}
for $i = 2,...,(m-1)$. This definition only plays a role at stationary points along the path, and it serves to smoothly switch between the two possible tangents $\boldsymbol{\tau}_i^+$ and $\boldsymbol{\tau}_i^-$ defined in equation \eqref{eq6}. Otherwise, when one image becomes higher in energy than another, the tangent changes abruptly, which can result in convergence issues. 

Finally, the tangent vector associated with each image needs to be normalized, namely
\begin{equation}\label{eq9}
    \hat{\boldsymbol{\tau}}_i = \frac{\boldsymbol{\tau}_i}{\lVert \boldsymbol{\tau}_i \rVert} \qquad \qquad i = 2,...,(m-1)
\end{equation}
With this modified tangent, the elastic band is well behaved and it converges reliably to the {\small MEP}, provided that a sufficient number of images are included in the path. In the cases of images with degenerate energies, the simple definition of the tangent as the normalized line segment connecting the previous and the next images to a given $i$-th image in the chain is recovered. Moreover, if the tangent vector has to be estimated also for the initial and the final fixed nuclear configurations, the previous definitions become ambiguous, and a specific choice has to be done. In these respects, all the technical aspects are reported in Section 2
of the supplementary material.

\subsection{Path minimization methods}\label{sec:path_minimization}

At each step of the {\small NEB} method, after the calculation of the tangent at each image, the total forces vector can be computed, by means of equation \eqref{eq3}, for each image along the path. At this point, the total forces vector has to be minimized for all the intermediate images. 
Two main minimization methods have been implemented in the \textsc{Crystal} code to find the {\small MEP}, starting from the initial nudged elastic band path: ($i$) the steepest descent (\textsc{Sd}), and ($ii$) the conjugate gradient (\textsc{Cg}) optimization methods. Details about the equations involved and their implementation are reported in Section 3
of the supplementary material, for both the path minimization methods. \\
To consider the minimization process converged to the minimum energy path, two criteria must be satisfied for each intermediate image, both associated with the real nuclear force vector orthogonal to the path:  
($i$) its Euclidean norm
and ($ii$) the root-mean-square of its Cartesian components must be less than given tolerances
\begin{equation}\label{eq10}
\begin{aligned}
    \text{($i$) } & \lVert \left.(\mathbf{F}_i)\right\vert_\perp \rVert < \Omega_{n} \\
    \text{($ii$) } & \text{\textsc{Rms}}\left[ \left.(\mathbf{F}_i)\right\vert_\perp \right] < \Omega_{r}
\end{aligned} \quad \qquad \forall \,\, i = 2,...,(m-1) \quad
\end{equation}
These requirements are based on the fact that, for an image located in the energy minimum, the real nuclear force component perpendicular to the path vanishes.
The default values for the two tolerances in \eqref{eq10} are $\Omega_n = 0.05$ eV/\AA$\,\approx$ 0.001 Ha/Bohr and $\Omega_r = \Omega_n/2$. 

\subsection{Beyond regular nudged elastic band method}

The main goal of the {\small NEB} method is the search for transition states, corresponding to saddle points in the {\small PES}. However, there are cases in which the resolution of the {\small MEP} near the saddle point is poor and the estimate of the activation energy obtained from the interpolation is subject to large uncertainty. Aiming at a better detection and characterization of the saddle points, two main variants of the basic {\small NEB} have been developed,\cite{henkelman_CIneb} as described below.

\subsubsection{Climbing Image method}

The climbing image approach constitutes a small modification to the basic {\small NEB} scheme.\cite{henkelman_CIneb} After a few iterations with the regular {\small NEB}, the image $i_{\text{max}}$ with the highest energy is identified. The force on this image is then computed using a modified version of equation \eqref{eq3}, given by
\begin{equation}\label{eq11}
\begin{aligned}
    \mathbf{F}_{i}^{\text{\tiny NEB}} & = \left.(\mathbf{F}_i)\right\vert_\perp \, - \, (\mathbf{F}_i)|_\parallel \\
    & = \mathbf{F}_i - 2(\mathbf{F}_i \cdot \hat{\boldsymbol{\tau}}_i) \, \hat{\boldsymbol{\tau}}_i
\end{aligned} \qquad \quad \text{ for } \quad i = i_{\text{max}} \,\,
\end{equation}
Equation \eqref{eq11} represents the full force due to the real potential, but with the component along the elastic band inverted.
By using this force expression for the maximum energy image, it becomes completely unaffected by the spring forces, so that the spacing between adjacent images will be different on each side of the climbing image. As it moves towards the saddle point, the images on one side will become compressed, while those on the other side will spread out. Qualitatively, the climbing image moves up the potential energy surface along the elastic band and down the potential surface perpendicular to the band. The only challenge remained at this point is ensuring there are enough images near the climbing image to obtain an accurate estimate of the reaction coordinate, since this determines the climbing direction. 

\subsubsection{Variable Spring Constants method}\label{sec:vark_method}

Since the saddle point is the most important point along the {\small MEP}, it is preferable to have more resolution in the neighborhood of the saddle point than near the endpoints. Indeed, as the images are moved closer to the saddle point, the approximation of the tangent will become more accurate. In such cases, it is more efficient to distribute the images unevenly along the path.
This can be accomplished by using higher values of the elastic constants near the saddle point.
On the base of Ref. \cite{henkelman_CIneb}, a scheme can be used where the spring constant depends linearly on the energy of the images, in such a way that images with low energy get connected by a weaker spring constant. This can be accomplished using the formula
\begin{equation}\label{eq12}
    k_i = \begin{cases}
    \displaystyle k_{\text{max}} - \Delta k \left(\frac{E_{\text{max}} - E_{ik}}{E_{\text{max}} - E_{\text{ref}}}\right) & \text{ if } \,\, E_{ik} > E_{\text{ref}} \\
    \displaystyle k_{\text{max}} - \Delta k = k_{\text{min}} & \,\, \text{if } \,\, E_{ik} \le E_{\text{ref}}
\end{cases}
\end{equation}    
with $\Delta k = k_{\text{max}} - k_{\text{min}}$ and for $i = 1,...,m$.
The energy $E_{ik}$ is defined as
\begin{equation}\label{eq13}
    E_{ik} = \begin{cases}
    \displaystyle \max\{E_i,E_{i+1}\} & \text{for} \quad i = 1,...,m-1 \\
    \displaystyle \max\{E_i,E_{i-1}\} & \text{for} \quad i = m
\end{cases}
\end{equation}    
and it represents the highest energy between the two images connected by the $i$-th spring, while $E_{\text{max}}$ is the maximum value of $E_i$ for the whole elastic band. Finally, $E_{\text{ref}}$ is a reference value for the energy, defining a minimum value of the spring constant. In Ref. \cite{henkelman_CIneb}, $E_{\text{ref}}$ is taken to be the energy of the highest energy endpoint in the path, that is
\begin{equation}\label{eq14}
    E_{\text{ref}} = \max\{E_{\text{\tiny IS}},E_{\text{\tiny FS}}\}
\end{equation}
where $E_{\text{\tiny IS}}$ and $E_{\text{\tiny FS}}$ are the energies of the initial and final states, respectively.
This choice ensures that the density of images is roughly equal near the two endpoints, even for highly asymmetric paths. The spring constant is, therefore, linearly scaled from a maximum value of $k_{\text{max}}$ for highest energy images to a minimum value of $k_{\text{max}} - \Delta k = k_{\text{min}}$ for the images with energy equal to $E_{\text{ref}}$ or lower. 

Using the variable spring constants method, the parallel component of the spring forces for each image is defined in a slightly different way with respect to equation \eqref{eq5}, namely
\begin{equation}\label{eq16}
 (\mathbf{F}_i^s)|_\parallel \, = \, \left(k_{i} \lVert\bm{q}_{i+1} - \bm{q}_i\rVert - k_{i-1} \lVert\bm{q}_{i} - \bm{q}_{i-1}\rVert\right) \hat{\boldsymbol{\tau}}_i 
\end{equation}
for $i = 2, ..., (m-1)$, so that a unique value of the elastic constant is assigned to each spring that describes the interaction among images. In equation \eqref{eq16}, the value of the elastic constant associated with the spring connecting the $i$-th and the $(i+1)$-th images only depends on the value $k_i$ of the elastic constant conferred to the $i$-th image, estimated by means of equation \eqref{eq12}.

\subsection{Minimum energy path interpolation}\label{sec:mep_interpolation}

The analysis of the results of a {\small NEB} calculation requires an interpolation between the images, in order to get estimates of the coordinates of atoms and the energy at maxima and minima along the {\small MEP}. For a good interpolation scheme, it is better to include the components of the real forces parallel to the path in the definition of the interpolation function, that can lead to the identification of extrema in the {\small MEP}, which could not be seen if only the energy data are taken into account.\cite{henkelman_new_tangent}

In the \textsc{Crystal} code, a cubic polynomial to represent the {\small MEP} between each pair of adjacent images is used. The polynomial describing the curve between two points (images) in the range $[\bm{q}_i, \bm{q}_{i+1}]$, being $\bm{q}_i$ the starting point (image) and $\bm{q}_{i+1}$ the final point (image) of the $i$-th line, can thus be written as
\begin{equation}\label{eq19}
    p(\bm{q}) = a_i \, \lVert \bm{q} - \bm{q}_i \rVert^3 + b_i \, \lVert \bm{q} - \bm{q}_i \rVert ^2 + c_i \, \lVert \bm{q} - \bm{q}_i \rVert + d_i 
\end{equation}
valid for all points $\bm{q} \in [\bm{q}_i, \bm{q}_{i+1}]$. The equations describing the coefficients of the polynomial in equation \eqref{eq19} can be derived by imposing four conditions for the matching of both the energy and the force at the two endpoints of the interval. That is, given a starting point $\bm{q}_i$ with ground state electronic energy $E_i \equiv E(\bm{q}_i)$ and real forces component parallel to the path provided by the projection
\begin{equation}\label{eq20}
    F_i \equiv \mathbf{F}_i \cdot \hat{\boldsymbol{\tau}}_i = -\nabla E(\bm{q}_i) \cdot \hat{\boldsymbol{\tau}}_i
\end{equation}
as well as a endpoint $\bm{q}_{i+1}$ with ground state electronic energy $E_{i+1} \equiv E(\bm{q}_{i+1})$ and component of the real forces parallel to the path defined following equation \eqref{eq20}, the four conditions are
\begin{equation}\label{eq21}
\begin{aligned}
    & p(\bm{q}_i) && \hspace{-0.288cm} = E_i \\[1.68ex] 
    & p(\bm{q}_{i+1}) && \hspace{-0.288cm} = E_{i+1}
\end{aligned}
\end{equation}
\vspace{-0.12cm}
\begin{equation}\label{eq22}
\begin{aligned}
    & \left. \frac{\partial p(\bm{q})}{\partial\lVert \bm{q} - \bm{q}_i\rVert} \right\rvert_{\bm{q} \, = \, \bm{q}_i} && \hspace{-0.4cm} = - F_i \\[1.68ex] 
    & \left. \frac{\partial p(\bm{q})}{\partial\lVert \bm{q} - \bm{q}_i\rVert} \right\rvert_{\bm{q} \, = \, \bm{q}_{i+1}} && \hspace{-0.4cm} = - F_{i+1}
\end{aligned}
\end{equation}
By imposing these conditions on the cubic polynomial \eqref{eq19}, the four parameters for the interpolation curve between two adjacent images have the form
\begin{align}
    a_i & = \frac{2(E_{i} - E_{i+1})}{\lVert \bm{q}_{i+1} - \bm{q}_i \rVert^3} - \frac{F_i + F_{i+1}}{{\lVert \bm{q}_{i+1} - \bm{q}_i \rVert}^2} \label{eq23} \\ 
    \notag \\
    b_i & = \frac{3(E_{i+1} - E_{i})}{\lVert \bm{q}_{i+1} - \bm{q}_i \rVert^2} + \frac{2F_i + F_{i+1}}{{\lVert \bm{q}_{i+1} - \bm{q}_i \rVert}} \label{eq24}
\end{align}
\vspace{-0.1cm}
\begin{align}
    c_i & = - F_i \\
    \notag \\
    d_i & = E_i
\end{align}

The cubic polynomial interpolation discussed above is typically quite smooth.\cite{henkelman_new_tangent} 
It is worth noting that the cubic interpolation introduced above does not guarantee the continuity of second derivatives at the images. However, adopting interpolation polynomial of higher orders, although ensuring continuous second derivatives, can add spurious oscillations between couples of images.\cite{henkelman_new_tangent}
Thus, if a continuous second derivative is not essential, the straightforward cubic interpolation outlined above is likely the preferred choice.

\section{Results and Discussion}\label{sec:results}

To assess the accuracy and the general applicability of the {\small NEB} implementation in the \textsc{Crystal} code, a set of test cases covering different chemical environments and reaction types were selected. These include elementary gas phase reactions and a more complex process, relevant in solid state chemistry.
In particular, we examined two molecular reactions, namely, the collinear proton transfer reaction,
\begin{equation}\label{eq:reaction1}\tag{R1}
	\text{H}_2 + \text{H} \rightarrow \text{H} + \text{H}_2
\end{equation}
and the keto-enol tautomerization of formamide, 
\begin{equation}\label{eq:reaction2}\tag{R2}
    \text{HC(=O)NH}_2 \rightarrow \text{HC(-OH)NH}
\end{equation}

These tests are discussed in Section \ref{sec:gas_phase_results} and provide a well-balanced set of systems in terms of complexity and chemical diversity, allowing for a comprehensive validation of the {\small NEB} algorithms in the molecular regime.
In addition, a bulk process involving a proton transfer between two oxygen atoms in a chabazite zeolite was simulated, to test the method in a fully periodic condensed phase environment. The results and the analysis of this process are discussed in Section \ref{sec:chabazite_results}.
Together, these tests aim to demonstrate the flexibility and reliability of the {\small NEB} algorithm in the \textsc{Crystal} code, across different length scales and chemical contexts.

\subsection{Gas Phase reactions}\label{sec:gas_phase_results}

\subsubsection{The collinear proton transfer reaction \eqref{eq:reaction1}}\label{sec:collinear_proton_tranfer_results}


The collinear proton transfer reaction \eqref{eq:reaction1} in the framework of Hartree-Fock and {\small DFT} has been extensively studied by B. S. Jursic,\cite{h2h_jursic_1997,h2h_jursic_1998} using \textsc{Gaussian 94} computational package.\cite{gaussian94} In particular, using a cc-p{\small VQZ} basis set, the activation barrier for the reaction \eqref{eq:reaction1} has found to be equal to 16.8 kcal/mol $\approx$ 0.7285 eV and 2.4 kcal/mol $\approx$ 0.1041 eV, respectively using Hartree-Fock Hamiltonian and {\small BLYP} exchange-correlation functional.\cite{h2h_jursic_1998} 
The experimental activation barrier used in literature to be compared with theoretical results has been computed by W. R. Schulz \emph{et al.}\cite{h2h_expt} from reaction rate versus temperature data over a temperature range of about $(330, 444)$ K,\cite{h2h_expt} for which the simple Arrhenius form was suitable and the reaction rates were available, and it has a value of 9.70 kcal/mol $\approx$ 0.4206 eV.\cite{h2h_expt} However, it has been observed\cite{h2h_temelso_2006} that the experimentally deduced activation barriers depend on the temperature range used for the Arrhenius fit, thus complicating a direct comparison with reaction barriers computed quantum mechanically.
Indeed, the previously mentioned ab initio calculations by B. S. Jursic,\cite{h2h_jursic_1997,h2h_jursic_1998} together with more recent theoretical results,\cite{h2h_temelso_2006,h2h_johnson_1994} confirm the finding that neither hybrid nor gradient-corrected density functionals methods are capable of reproducing the experimental reaction barrier found by W. R. Schulz \emph{et al.},\cite{h2h_expt} and only a coupled-cluster theory with single, double, and perturbative triple
substitutions [{\small CCSD(T)}] approach,\cite{ccsdt_method_1989} can reach activation energy values comparable to the experimental one.\cite{h2h_temelso_2006,h2h_johnson_1994}

In this work, avoiding the debated and controversial comparison with experimental data, for the sake of testing our implementation against other theoretical methods and quantum mechanical packages, we employed unrestricted Hartree-Fock ({\small UHF}) Hamiltonian, using an all electron basis set consisting of 6 atomic orbitals for each hydrogen atom,\cite{H_5-11G*_dovesi_1984} and we compare our results with those obtained using the \textsc{Orca} code,\cite{orca_code} within the same computational setup and {\small NEB} methods.
The total number of images along the path is taken equal to $m = 9$, including the initial and final states of the reaction. The initial and final states of the reaction are shifted with respect to the corresponding center of mass of the system, before starting the {\small NEB} calculation. The nuclear positions for the initial, final and intermediate states at each {\small NEB} step for all the calculations listed in Table \ref{tab:h2h_results} are made available through a Zenodo repository.\cite{zenodo_neb}
The initial reaction path is obtained through linear interpolation between the Cartesian nuclear coordinates of the atoms in the initial and final states, while the minimization of the {\small NEB} path towards the {\small MEP} is achieved using two algorithms, namely, steepest descent (\textsc{Sd}) method, with an optimization step equal to $ds = 1.0$ Bohr$^2$/Ha, and the conjugate gradient (\textsc{Cg}) technique, with $ds = 0.8$ Bohr$^2$/Ha. The calibration of the optimization step for the path minimization methods is detailed in Section 8
and Table 3
of the supplementary material.

The results obtained using the different {\small NEB} techniques, in conjunction with the two procedures for path optimization, are listed in Table \ref{tab:h2h_results}, together with the corresponding results computed with the \textsc{Orca} code.\cite{orca_code}
The energy barriers, at the {\small UHF} level, are in striking agreement with the ones found by the \textsc{Orca} code, as well as with the other theoretical references. For all the calculations, the vibrational frequencies of the transition state structure have one imaginary eigenvalue, whose value is reported in Table \ref{tab:h2h_results}, thus confirming the detection of a saddle point in the {\small PES}.

\begin{table}[htb]
	\renewcommand{\arraystretch}{1.2}
	\setlength{\tabcolsep}{4pt}
	\begin{tabular}{ll|c|c|c}
		\multicolumn{5}{l}{Reaction \eqref{eq:reaction1} with $m = 9$ images - {\footnotesize UHF} method} \\
		\toprule
		\multicolumn{2}{c|}{Method} & $n_{opt}$ & $E_a$ [eV] & $\nu$ [cm$^{-1}$] \\
		\toprule
		\multirow{3}{*}{Basic {\footnotesize NEB} $\,\,\,$} & \textsc{Sd} & 58 & 0.721139 & 2267.5720$\,i$ \\
		& \textsc{Cg} & 76 & 0.721134 & 2266.9278$\,i$ \\
		\cline{2-5}
		& \textsc{Orca}-\textsc{Lbfgs} & 20 & 0.721145 & 2231.0023$\,i$ \\
		\hline
		\multirow{3}{*}{{\footnotesize CI-NEB}} & \textsc{Sd} & 58 & 0.721139 & 2267.5719$\,i$ \\
		& \textsc{Cg} & 76 & 0.721134 & 2266.9285$\,i$ \\
		\cline{2-5}
		& \textsc{Orca}-\textsc{Lbfgs} & 20 & 0.721145 & 2246.6890$\,i$ \\
		\hline
		\multirow{3}{*}{{\footnotesize VARK-NEB}} & \textsc{Sd} & 58 & 0.721139 & 2267.5727$\,i$ \\
		& \textsc{Cg} & 76 & 0.721134 & 2266.9284$\,i$ \\
		\cline{2-5}
		& \textsc{Orca}-\textsc{Lbfgs} & 27 & 0.721145 & 2217.8591$\,i$ \\
		\toprule
		\multicolumn{3}{c|}{B. S. Jursic\cite{h2h_jursic_1998}} & 0.72852 & 2476$\,i$ \\
		\multicolumn{3}{c|}{B. G. Johnson \emph{et al.}\cite{h2h_johnson_1994}} & 0.721145 & 2296$\,i$ \\
		\toprule
	\end{tabular}
    \caption{\small Number of optimization steps $n_{opt}$ required to find the {\footnotesize MEP}, activation energy $E_a$ and imaginary frequency $\nu$ of the transition state. Different {\footnotesize NEB} methods (see Section \ref{sec:theory}) and various optimization schemes (see Section \ref{sec:path_minimization}) are compared with the results obtained with the \textsc{Orca} code,\cite{orca_code} adopting the same computational setup. The optimization steps in the \textsc{Sd} and \textsc{Cg} path minimization algorithms are reported in the main text of Section \ref{sec:collinear_proton_tranfer_results}.}\label{tab:h2h_results}
\end{table}

\begin{figure*}[htb]
	\includegraphics[scale=0.392]{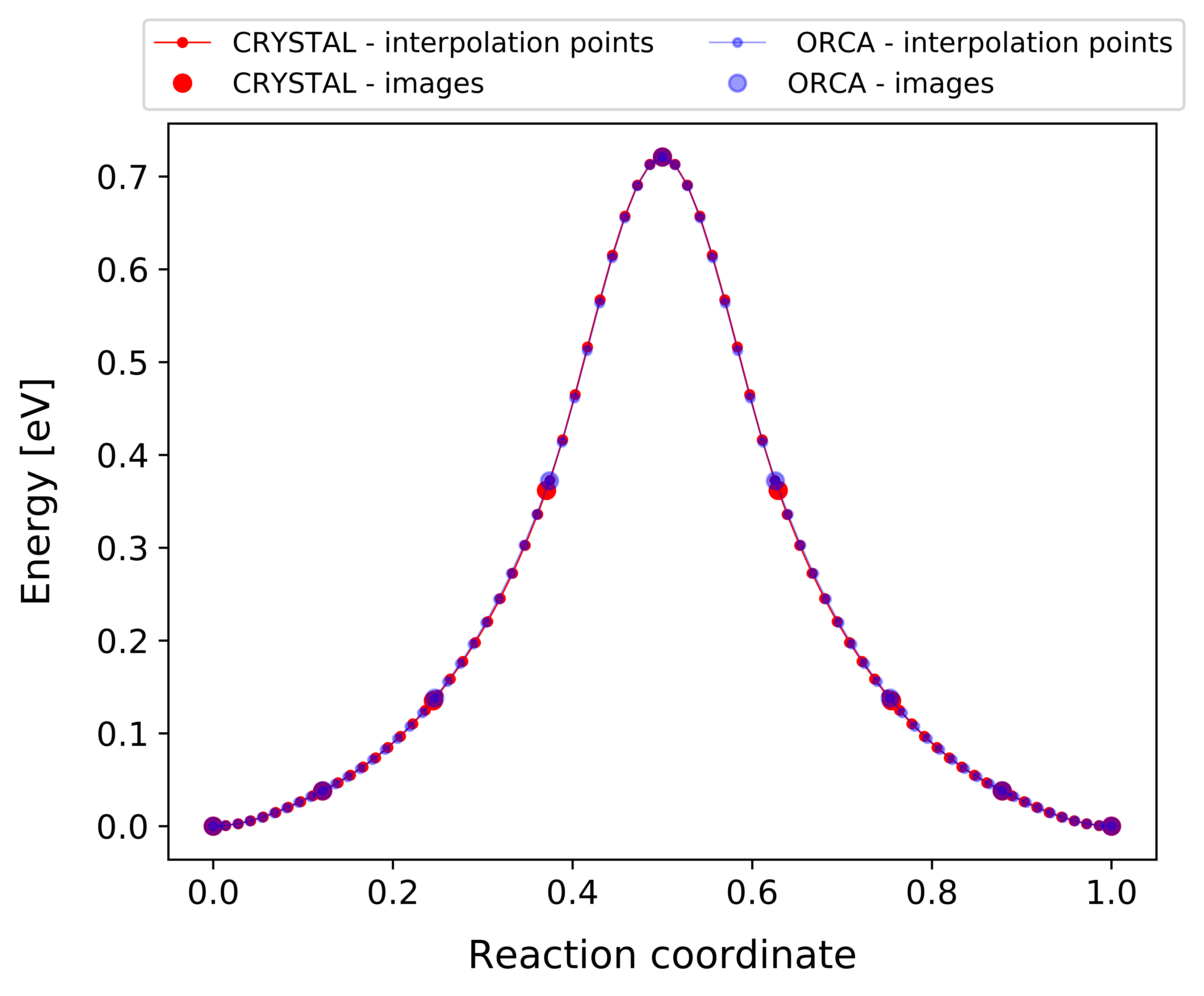} \hspace{0.05cm} \includegraphics[scale=0.392]{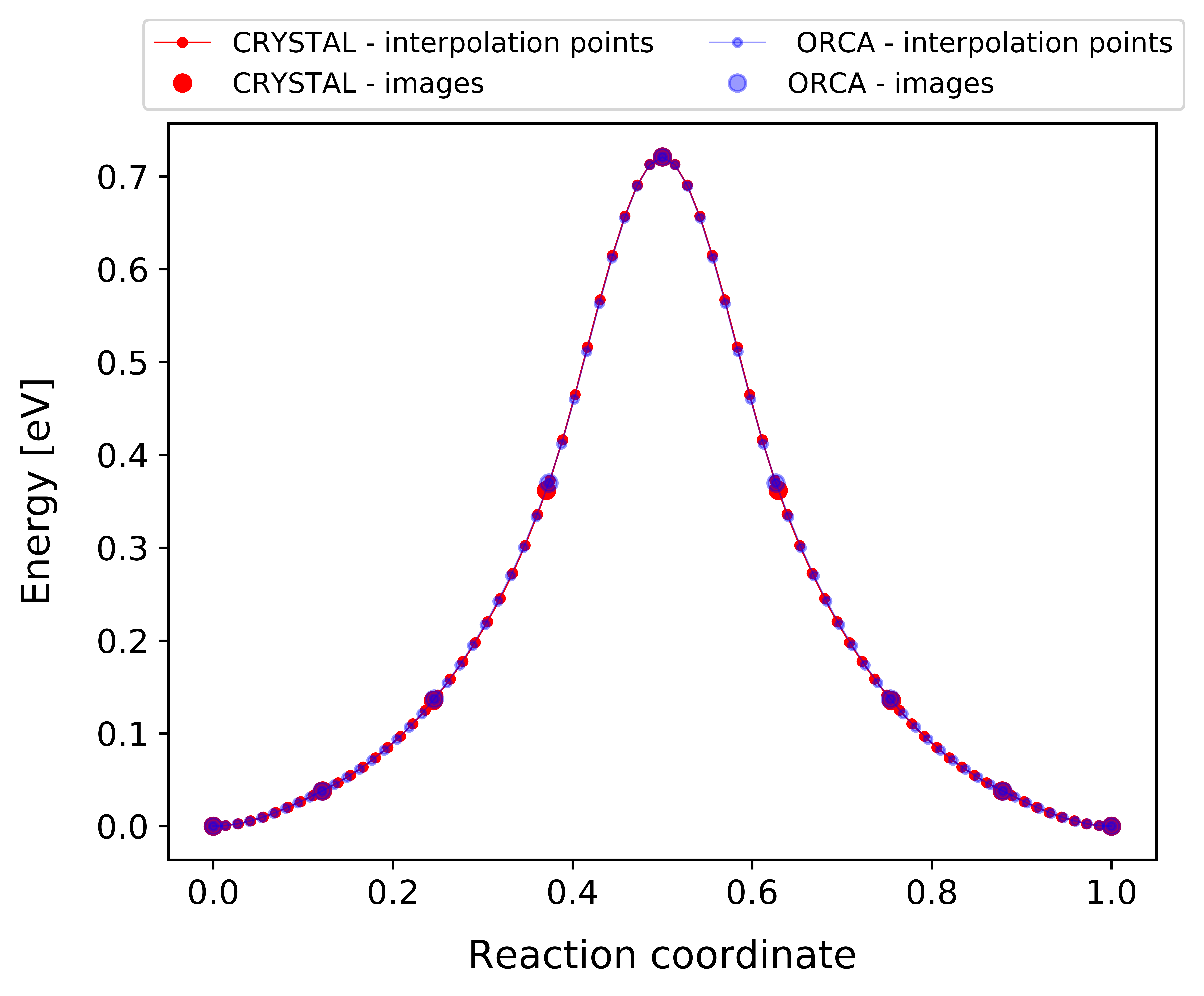} \hspace{0.05cm}
	\includegraphics[scale=0.392]{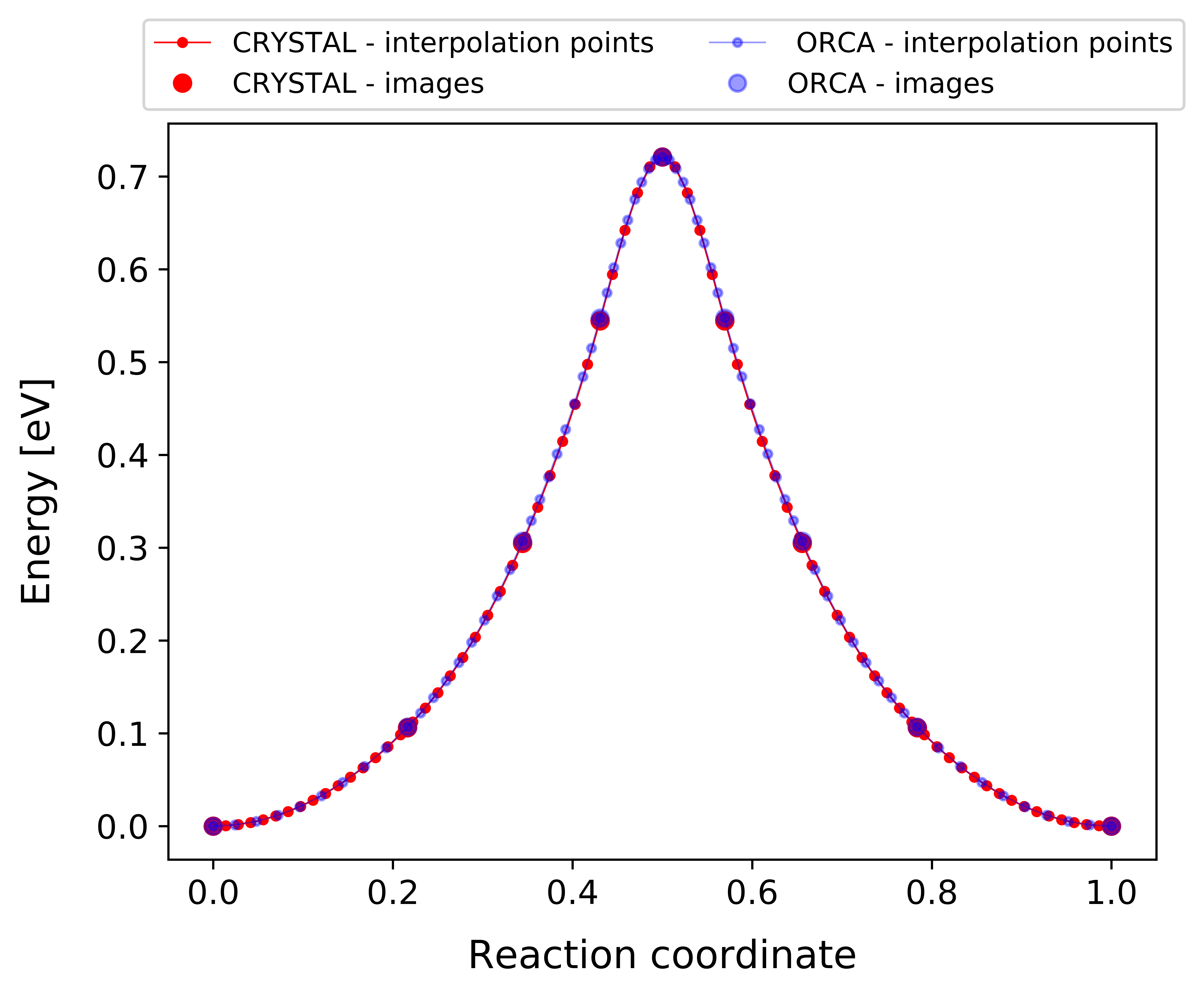}
	\caption{\small Minimum energy paths obtained with the basic {\footnotesize NEB} (left panel), the climbing image ({\footnotesize CI-NEB}, central panel) and the variable spring constants ({\footnotesize VARK-NEB}, right panel) methods, using the steepest descent (\textsc{Sd}) algorithm with an optimization step of $ds = 1.0$ Bohr$^2$/Ha to converge the reaction path towards the minimum energy path. Details about the activation barriers, as well as about the vibrational frequencies for the transition state configurations, are reported in Table \ref{tab:h2h_results}. The interpolation of the minimum energy paths is done using 72 points. The method used for the interpolation is described in Section \ref{sec:mep_interpolation}.}\label{fig:h2h_reaction_paths}
\end{figure*}

Examples of the symmetric form of the {\small MEP} computed with the basic, climbing image ({\small CI-NEB}) and variable spring constants ({\small VARK-NEB}) methods, using the steepest descent minimization algorithm, are reported in Figure \ref{fig:h2h_reaction_paths}, superimposed to the corresponding minimum energy paths obtained using the same computational setup within the \textsc{Orca} code, showing a perfect agreement between the results of the two programs.

\subsubsection{The keto-enol tautomerization of formamide \eqref{eq:reaction2}}\label{sec:formamide_results}

The tautomerization reaction \eqref{eq:reaction2}, describing the proton transfer from the hydroxyl group to the amino nitrogen in formimidic acid, producing the amide form of formamide, has been extensively investigated in literature by several first principles electronic structure studies. Among these, the pioneering work by Schlegel \emph{et al.}\cite{schlegel_1982} already highlighted the energetic preference of the amide form over the enolic tautomer, with an estimated tautomerization enthalpy of about 12 kcal/mol $\approx$ 0.52 eV, including zero-point energy corrections and {\small MP2}-level electron correlation. Later, more refined benchmark calculations conducted by G. Fogarasi \cite{fogarasi_2010} confirmed this result, reporting a best estimate energy difference of about 10.6 kcal/mol (0.46 eV) between the two forms at the {\small CCSD(T)//MP2/}aug-cc-p{\small VTZ} level of theory.
C. Adamo and coworkers\cite{adamo_1997} further analyzed this reaction using {\small B3LYP} functional and post-Hartree-Fock methods, finding a barrier for direct intramolecular proton transfer in gas phase equal to 46.2 kcal/mol $\approx$ 2.00 eV when computed at the {\small B3LYP/6-31G}(d,p) level, the same value also reported by G. Fogarasi\cite{fogarasi_2010} using the same level of theory. 
	
In the present work, we investigate reaction \eqref{eq:reaction2} using the {\small B3LYP} exchange-correlation functional and an all-electron basis set consisting of 57 atomic orbitals (see Section \ref{sec:computational_details} for further details). The number of images for the {\small NEB} calculations was set equal to $m = 9$, including the initial and final states of the reaction.
The minimization of the {\small NEB} path towards the {\small MEP} is computed using two different algorithms, namely, the steepest descent (\textsc{Sd}) method, with an optimization step equal to $ds = 1.0$ Bohr$^2$/Ha, and the conjugate gradient (\textsc{Cg}) technique, with $ds = 0.8$ Bohr$^2$/Ha. For a detailed discussion on the selection of the optimization step, refer to Section 8
and Table 4
of the supplementary material.

The nuclear positions for the initial, final and intermediate states at each {\small NEB} step for all the calculations listed in Table \ref{tab:formamide_results} are made available through a Zenodo repository.\cite{zenodo_neb}

The bond lengths of the optimized structure for the reactant and the product of the reaction \eqref{eq:reaction2}, used as initial and final states in the {\small NEB} calculations, as well as the bond lengths of the transition states obtained through the {\small NEB} methods listed in Table \ref{tab:formamide_results}, are reported in Table 2
of the supplementary material, together with the experimental values\cite{brown_1987} for the initial and final geometric structures, and the theoretical results obtained by C. Adamo and coworkers.\cite{adamo_1997}

\begin{figure}[htb]
	\includegraphics[scale=0.288]{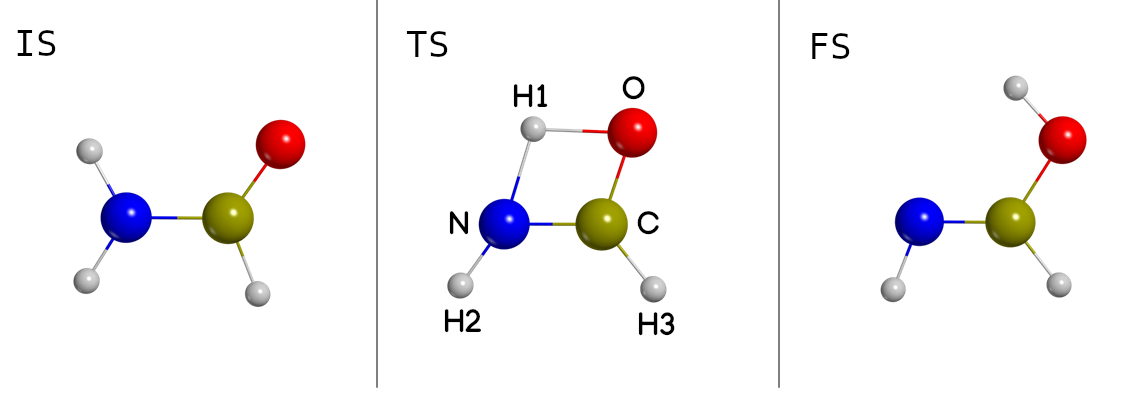}
	\caption{\small Initial ({\footnotesize IS}), transition ({\footnotesize TS}) and final ({\footnotesize FS}) states for the keto-enol tautomerization of formamide reaction. Dark olive, blue, red and white spheres represent carbon, nitrogen, oxygen and hydrogen atoms, respectively. The transition state is computed using the {\footnotesize CI-NEB} method in conjunction with the \textsc{Sd} path minimization scheme. The associated {\footnotesize MEP} is reported in the central panel of Figure \ref{fig:formamide_reaction_paths}.
    }\label{fig:formamide}
\end{figure}

A picture of the initial, final, and transition state configurations, the latter obtained by means of the {\small CI-NEB} method in conjunction with the steepest descent path minimization, is reported in Figure \ref{fig:formamide}.
	
\begin{figure*}[htb]
	\includegraphics[scale=0.384]{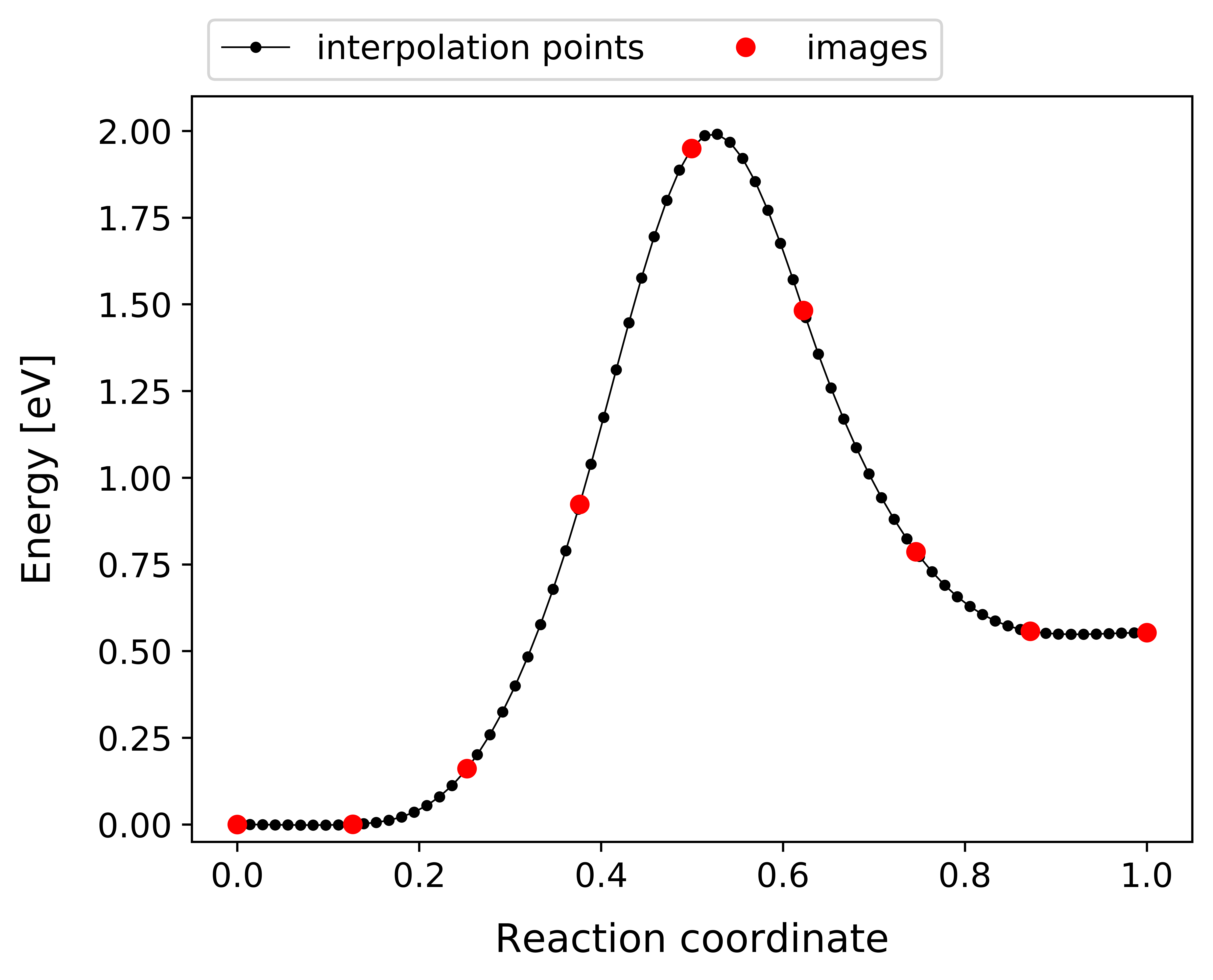} \hspace{0.2cm}
	\includegraphics[scale=0.384]{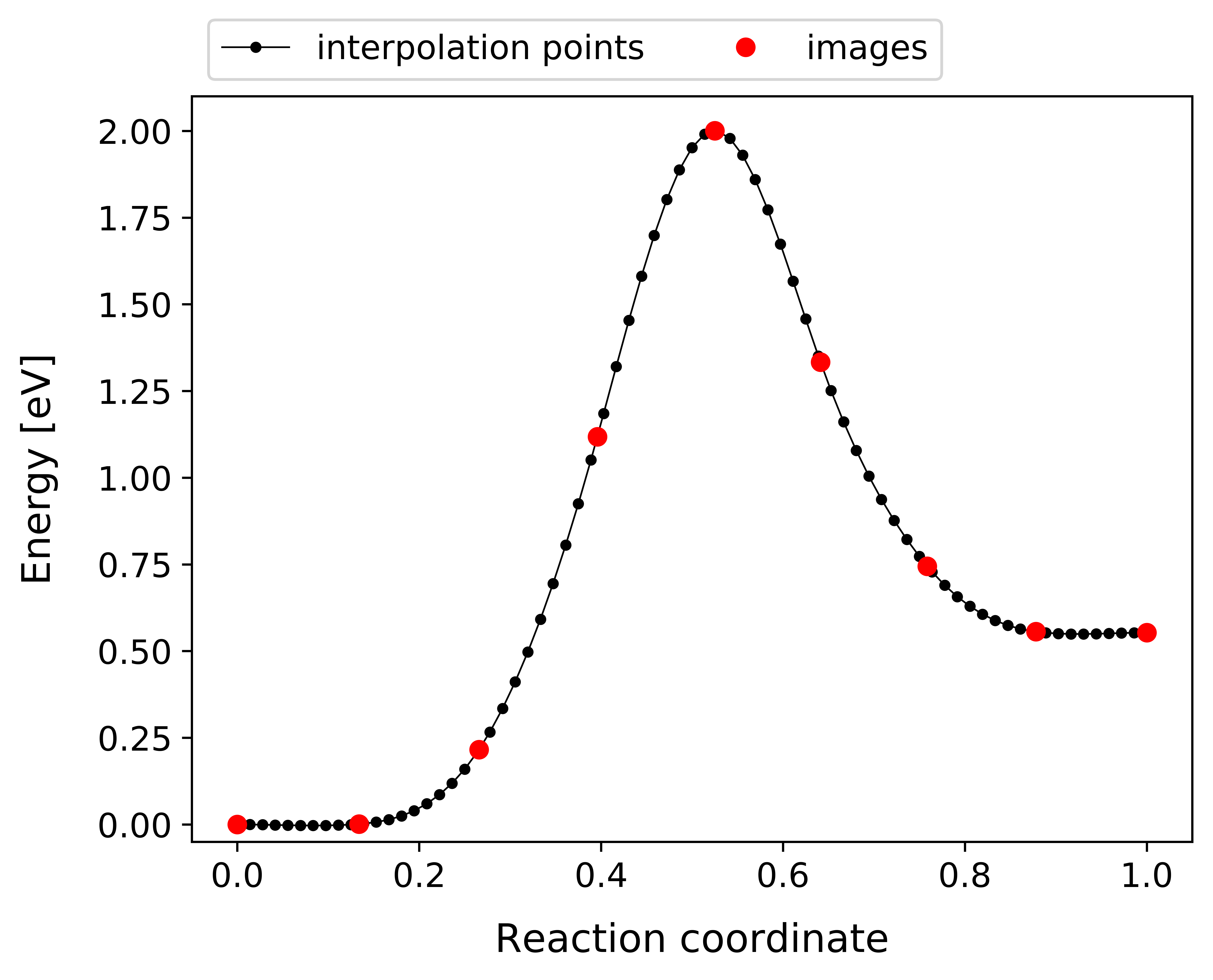} \hspace{0.2cm}
	\includegraphics[scale=0.384]{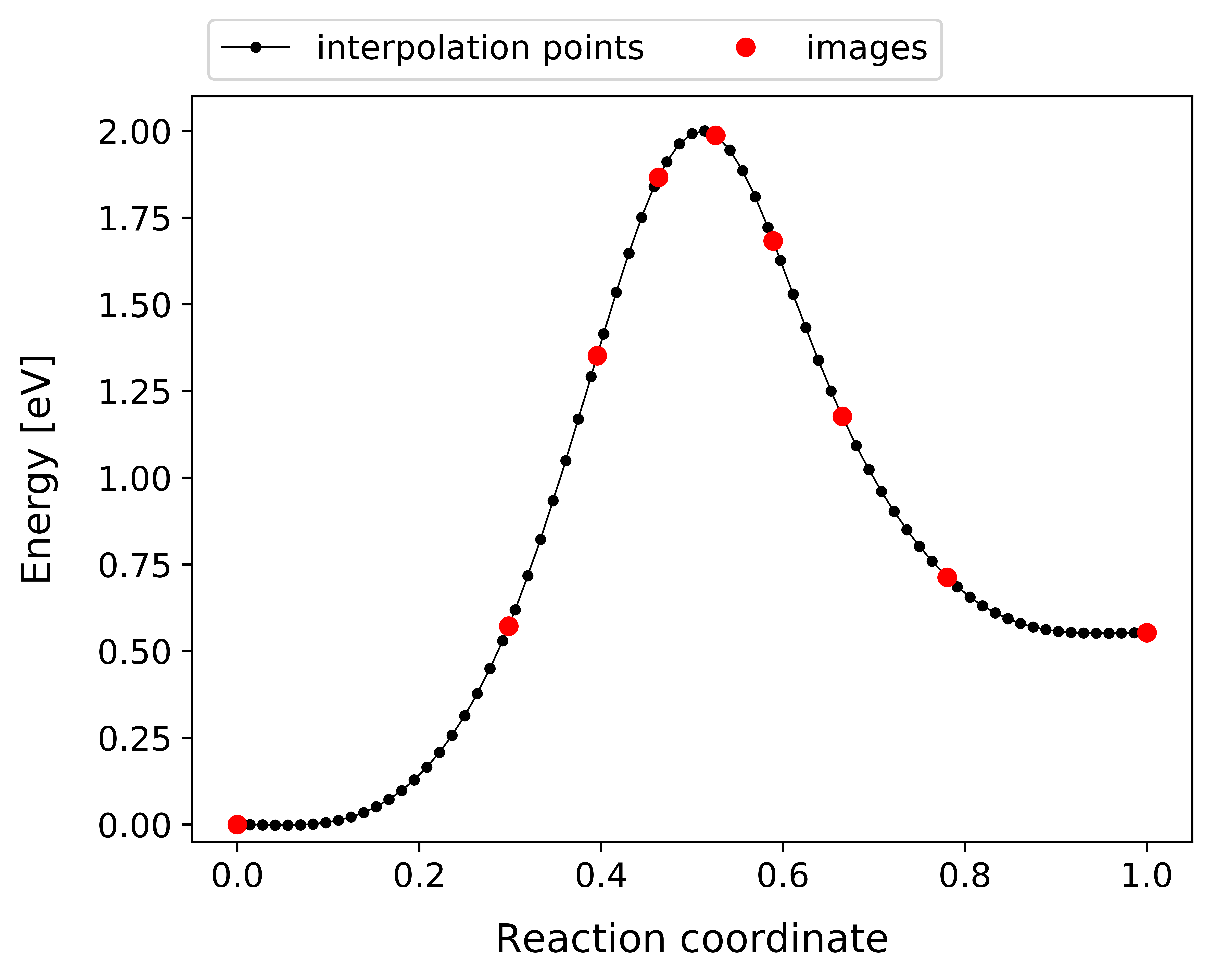}
	\caption{\small Minimum energy paths obtained with the basic {\footnotesize NEB} (left panel), the climbing image ({\footnotesize CI-NEB}, central panel) and the variable spring constants ({\footnotesize VARK-NEB}, right panel) methods, using the steepest descent (\textsc{Sd}) algorithm with an optimization step of $ds = 1.0$ Bohr$^2$/Ha to converge the reaction path towards the minimum energy path. Details about the activation barriers, as well as about the vibrational frequencies for the transition state configurations, are reported in Table \ref{tab:formamide_results}. The interpolation of the minimum energy paths is done using 72 points. The method used for the interpolation is described in Section \ref{sec:mep_interpolation}.}\label{fig:formamide_reaction_paths}
\end{figure*}

The computed activation barriers, summarized in Table \ref{tab:formamide_results}, range from 1.95 eV to 2.00 eV (depending on the {\small NEB} method and the reaction path minimization scheme employed), in excellent agreement with the theoretical value of 
46.2 kcal/mol $\approx$ 2.00 eV reported by C. Adamo \emph{et al.}\cite{adamo_1997} as well as by G. Fogarasi,\cite{fogarasi_2010} obtained using {\small B3LYP} functional in conjunction with the {\small 6-31G}(d,p) basis set. Furthermore, the calculated tautomerization energy obtained from the energy difference between the reactant (formamide) and product (formamidic acid) states amounts to $\Delta E_r \approx 0.55$ eV, in good agreement with the best estimates from previous studies.\cite{schlegel_1982,fogarasi_2010}	
The transition state has been characterized by a vibrational frequency analysis, identifying the presence of a single imaginary eigenvalue, with values (reported in Table \ref{tab:formamide_results}) consistent with a proton transfer phenomenon, thus confirming the identification of a first order saddle point. Moreover, our calculated values are in good agreement with the imaginary frequency of 1894$\,i$ cm$^{-1}$ reported by G. Fogarasi,\cite{fogarasi_2010} calculated at the {\small MP2/}cc-p{\small VTZ} level of theory.
The {\small MEP}s obtained using the three {\small NEB} methods (i.e., basic, {\small CI-NEB} and {\small VARK-NEB}), in conjunction with the steepest descent path optimization, are reported in Figure \ref{fig:formamide_reaction_paths}.
	
As in the collinear proton transfer case of Section \ref{sec:collinear_proton_tranfer_results}, we observe that the three {\small NEB} variants, combined with the selected minimization schemes, lead to consistent results, and all the transition state geometries exhibit a well-defined imaginary frequency associated with the reaction coordinate. 

\begin{table}[htb]
	\renewcommand{\arraystretch}{1.2}
	\setlength{\tabcolsep}{4pt}
	\begin{tabular}{ll|c|c|c}
		\multicolumn{5}{l}{Reaction \eqref{eq:reaction2} with $m = 9$ images - {\footnotesize B3LYP} functional} \\
		\toprule
		\multicolumn{2}{c|}{Method} & $n_{opt}$ & $E_a$ [eV] & $\nu$ [cm$^{-1}$] \\
		\toprule
		\multirow{2}{*}{Basic {\footnotesize NEB} $\,\,\,$} & \textsc{Sd} & 117 & 1.949339 & 1825.8207$\,i$ \\
		& \textsc{Cg} & 149 & 1.949700 & 1826.4288$\,i$ \\
		\hline
		\multirow{2}{*}{{\footnotesize CI-NEB}} & \textsc{Sd} & 111 & 2.000514 & 1905.1026$\,i$ \\
		& \textsc{Cg} & 143 & 2.000514 & 1905.1072$\,i$ \\
		\hline
		\multirow{2}{*}{{\footnotesize VARK-NEB}} & \textsc{Sd} & 75 & 1.986986 & 1870.2894$\,i$ \\
		& \textsc{Cg} & 98 & 1.986956 & 1870.1877$\,i$ \\
		\toprule
		\multicolumn{3}{c|}{C. Adamo \emph{et al.}\cite{adamo_1997}} & 2.00 & -- \\
		\multicolumn{3}{c|}{G. Fogarasi\cite{fogarasi_2010}} & 2.00 & -- \\
		\toprule
	\end{tabular}
	\caption{\small Number of optimization steps $n_{opt}$ required to find the {\footnotesize MEP}, activation energy $E_a$ and imaginary frequency $\nu$ of the transition state, resultant from different {\footnotesize NEB} methods (see Section \ref{sec:theory}) and various optimization schemes (see Section \ref{sec:path_minimization}) for the search of the minimum energy path. The optimization steps in the \textsc{Sd} and \textsc{Cg} path minimization algorithms are reported in the main text of Section \ref{sec:formamide_results}. }\label{tab:formamide_results}
\end{table}

\subsection{Proton exchange process in chabazite bulk}\label{sec:chabazite_results}

As a validation test for the implementation of the {\small NEB} method in the \textsc{Crystal} code in three-dimensional periodic systems, the proton jump between two Br\o nsted sites in the acidic chabazite was analyzed.

\begin{figure}[htb]
	\includegraphics[scale=0.288]{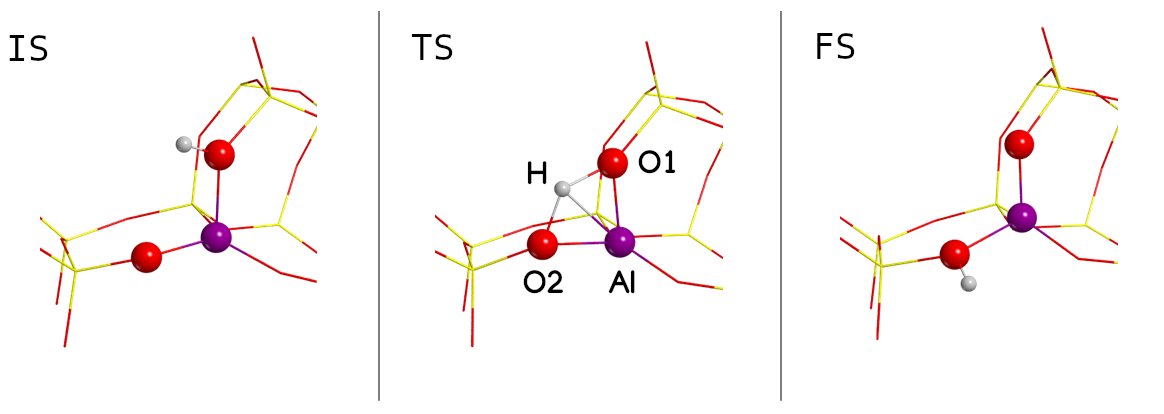}
	\caption{\small Initial ({\footnotesize IS}), transition ({\footnotesize TS}) and final ({\footnotesize FS}) states for the proton exchange process in the acidic chabazite bulk. Yellow, purple, red and white spheres represent silicon, aluminum, oxygen and hydrogen atoms, respectively. The transition state is computed using the {\footnotesize CI-NEB} method in conjunction with the \textsc{Sd} path minimization scheme, as implemented in the \textsc{Crystal} code. The associated {\footnotesize MEP} is reported in the central panel of Figure \ref{fig:chabazite_reaction_paths}. Details about the computational setup are given in Section \ref{sec:chabazite_results} of the main text.}\label{fig:chabazite}
\end{figure}

The acidic chabazite zeolite consists of a network of double six-membered alumino-silicate rings, connected by four-membered rings. 
The unit cell contains 37 atoms, with formula unit HAlSi$_{11}$O$_{24}$.
This system is widely studied in the literature as an exemplary case of reactivity in zeolites. 
M. Sierka \emph{et al.}\cite{chab_sierka_2001} have studied the proton jump in several proton-exchanged alumino-silicates, including chabazite. In their work, periodic calculations were carried out adopting the {\small QM-P}ot approach, in which the {\small QM} region, (involving the reaction site) was computed at a {\small B3LYP} level combining an Ahlrich's Gaussian double$\,$-$\,\zeta$ polarized (for H, Si and Al atoms) with a triple$\,$-$\,\zeta$ polarized (for O atoms) basis set, whereas the long range contribution was handled by molecular mechanics embedding.\cite{chab_sierka_2001} The energy barrier computed in this way, related to the proton transfer between oxygen sites O1 and O2 in dry chabazite (see Figure \ref{fig:chabazite}) is equal to 17.6 kcal/mol $\approx$ 73.6 kJ/mol.\cite{chab_sierka_2001}
At the same time, A. Rimola and coworkers\cite{chab_rimola_2010} have implemented, in the \textsc{Crystal} code, the distinguished reaction coordinate ({\small DRC}) method for the location of the transition states, using a set of redundant internal valence coordinates, with a localization of the saddle point based on the calculation of the Hessian matrix of the initial guess structure.\cite{chab_rimola_2010} For the proton-exchanged process in the acidic chabazite zeolite bulk, modeled with {\small B3LYP} exchange-correlation functional, they obtained an energy barrier of 15.0 kcal/mol $\approx$ 62.8 kJ/mol using the basic {\small DRC} method, a value of 18.3 kcal/mol $\approx$ 76.6 kJ/mol adopting the {\small DRC} method with fixed cell approach, and of 17.1 kcal/mol $\approx$ 71.5 kJ/mol using the basic {\small DRC} method with a triple$\,$-$\,\zeta$ polarized basis set.\cite{chab_rimola_2010}

In order to analyze the same proton-exchange process using the {\small NEB} method implemented in the \textsc{Crystal} code, a path with $m = 9$ images is considered, including initial and final states. The values of the elastic constants used for the basic, {\small CI-NEB} and {\small VARK-NEB} methods are reported in Section \ref{sec:computational_details}, together with other details about the {\small NEB} calculation setup.
The geometry of the initial image was fully optimized, allowing relaxation of both atomic positions and lattice parameters, in the absence of symmetry constraints. On the contrary, the structure of the final image has been optimized keeping the volume and the lattice parameters constant and equal to those obtained by optimizing the initial image.
This corresponds to a fixed cell approach, as all the images have the same reference cell, with $a = 9.4445$ \AA, $b = 9.3940$ \AA, $c = 9.3711$ \AA, $\alpha = 93.83^\circ$, $\beta = 94.47^\circ$, $\gamma = 94.79^\circ$.
The search of the {\small MEP} was performed using two minimization algorithms, namely, the steepest descent (\textsc{Sd}) and the conjugate gradient (\textsc{Cg}), both with a constant optimization step of $ds = 0.8$ Bohr$^2$/Ha.

The initial, final and the transition state geometrical configurations in the surrounding of the proton exchange center, obtained using the {\small CI-NEB} method in conjunction with the steepest descent path minimization scheme, are reported in Figure \ref{fig:chabazite}.
The nuclear positions for the initial, final and intermediate states at each {\small NEB} step for all the calculations listed in Table \ref{tab:chabazite_de_nu} are made available through a Zenodo repository.\cite{zenodo_neb}
The values of the distances between the atoms involved in the proton exchange process for the initial and final states are reported in Table \ref{tab:chabazite_distances}, while the same quantities for the transition state configuration, obtained with the three different {\small NEB} techniques, are listed in Table \ref{tab:chabazite_distances_ts}, together with the corresponding values of other theoretical works previously mentioned. The results for the activation barriers, as well as the imaginary frequencies associated with the transition state found using different {\small NEB} schemes within our current implementation, together with the theoretical literature values, are reported in Table \ref{tab:chabazite_de_nu}.

\begin{table}[htb]
	\renewcommand{\arraystretch}{1.2}
	\setlength{\tabcolsep}{4pt}
	\begin{tabular}{c|llll}
		\toprule
		Configuration & $d$(O1-H) & $d$(O2-H) & $d$(O1-Al) & $d$(O2-Al) \\
		\toprule
		\multirow{2}{*}{\footnotesize IS} & 0.9694 & 2.5645 & 1.8988 & 1.7028 \\
		& 0.969 [$a$] & 2.565 [$a$] & 1.899 [$a$] & 1.703 [$a$] \\
		\toprule
		\multirow{2}{*}{\footnotesize FS} & 3.3976 & 0.9713 & 1.7062 & 1.8801 \\
		& 3.398 [$a$] & 0.971 [$a$] & 1.877 [$a$] & 1.705 [$a$] \\
		\toprule
    \end{tabular}
    \caption{\small Distances [\AA] between the relevant atoms involved in the proton exchange process in the acidic chabazite, in the initial ({\footnotesize IS}) and final ({\footnotesize FS}) structures used as first and last images for the nudged elastic band calculations. Results labeled with [$a$] are taken from A. Rimola and coworkers.\cite{chab_rimola_2010} The labels of the atoms refer to Figure \ref{fig:chabazite}. All the calculations reported are performed using the hybrid {\footnotesize B3LYP} exchange-correlation functional.}\label{tab:chabazite_distances}
\end{table}

\begin{table}[htb]
	\renewcommand{\arraystretch}{1.2}
	\setlength{\tabcolsep}{1.48pt}
	\begin{tabular}{ll|cccc}
		\multicolumn{6}{l}{Transition state ({\footnotesize TS}) configuration} \\
		\multicolumn{6}{l}{Number of images $m = 9$ - {\footnotesize B3LYP} functional} \\
		\toprule
		\multicolumn{2}{c|}{Method} & $d$(O1-H) & $d$(O2-H) & $d$(O1-Al) & $d$(O2-Al) \\
		\toprule
		\multirow{2}{*}{Basic {\footnotesize NEB}} & \textsc{Sd} & 1.6115 & 1.0399 & 1.7679 & 1.8673 \\
		& \textsc{Cg} & 1.6113 & 1.0399 & 1.7679 & 1.8673 \\
		\hline
		\multirow{2}{*}{{\footnotesize CI-NEB}} & \textsc{Sd} & 1.2316 & 1.2647 & 1.8280 & 1.8193 \\
		& \textsc{Cg} & 1.2316 & 1.2647 & 1.8280 & 1.8193 \\
		\hline
		\multirow{2}{*}{{\footnotesize VARK-NEB}$\,\,$} & \textsc{Sd} & 1.2113 & 1.2876 & 1.8314 & 1.8158 \\
		& \textsc{Cg} & 1.2115 & 1.2874 & 1.8313 & 1.8158 \\
		\toprule
		\multicolumn{2}{l|}{\footnotesize DRC} & $\,\,\,\,$ 1.222 [$a$] & $\,\,\,\,$ 1.245 [$a$] & $\,\,\,\,$ 1.830 [$a$] & $\,\,\,\,$ 1.820 [$a$] \\
		\toprule
	\end{tabular}
	\caption{\small Distances [\AA] between the relevant atoms involved in the proton exchange process for the transition state configuration of the acidic chabazite, resultant from different \textsc{Neb} methods (see Section \ref{sec:theory}) and various optimization schemes (see Section \ref{sec:path_minimization}) for the search of the minimum energy path, compared with other theoretical results. Results labeled with [$a$] are taken from Rimola and coworkers.\cite{chab_rimola_2010} The labels of the atoms refer to Figure \ref{fig:chabazite}.}\label{tab:chabazite_distances_ts}
\end{table}

While the basic {\small NEB} method is not capable of reproducing the bond lengths nor the activation barriers found in other theoretical studies, both the {\small CI-NEB} and the {\small VARK-NEB} approaches lead to values in very good agreement with the existing literature.

\begin{figure*}[htb]
	\includegraphics[scale=0.39]{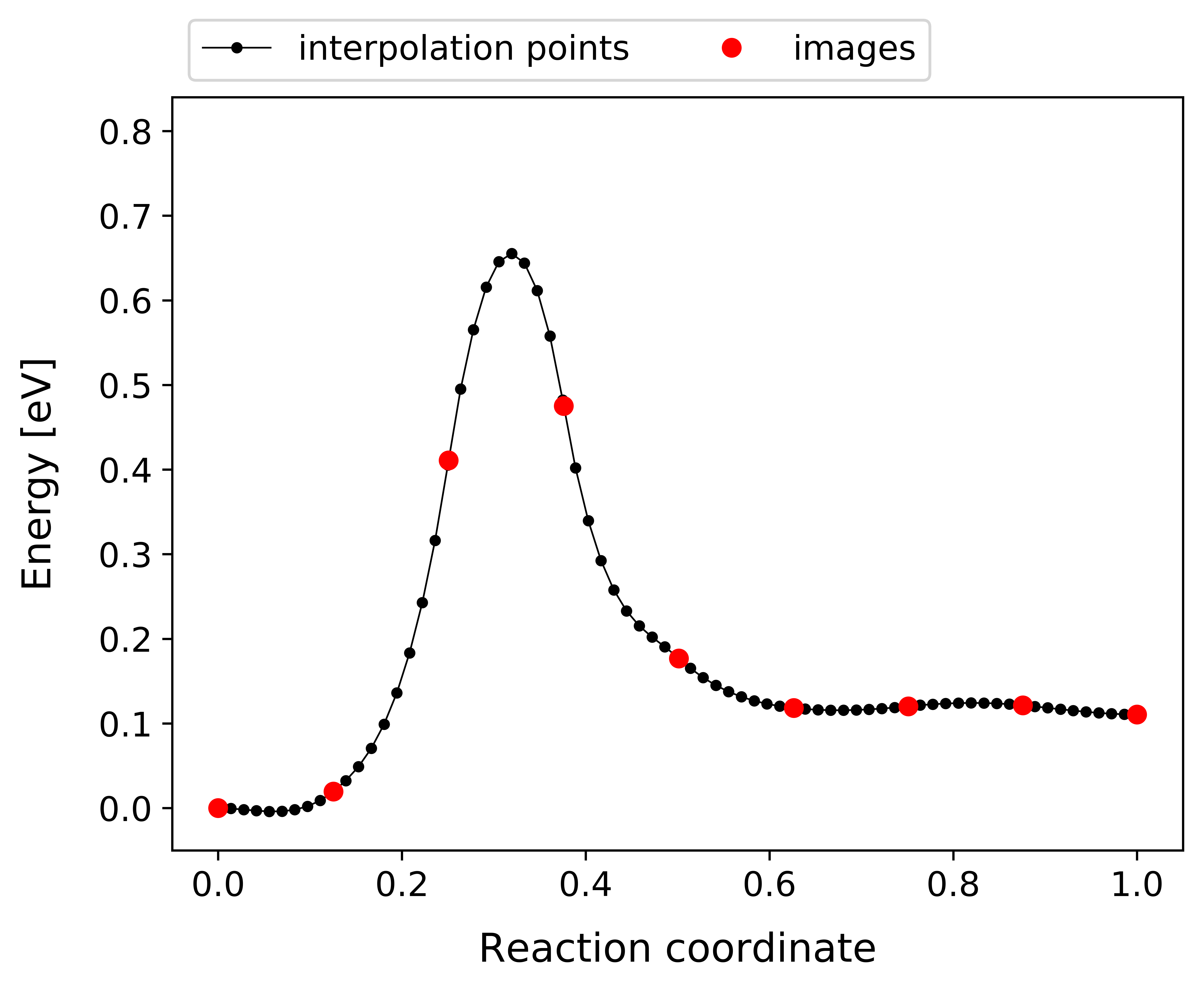} \hspace{0.2cm}
	\includegraphics[scale=0.39]{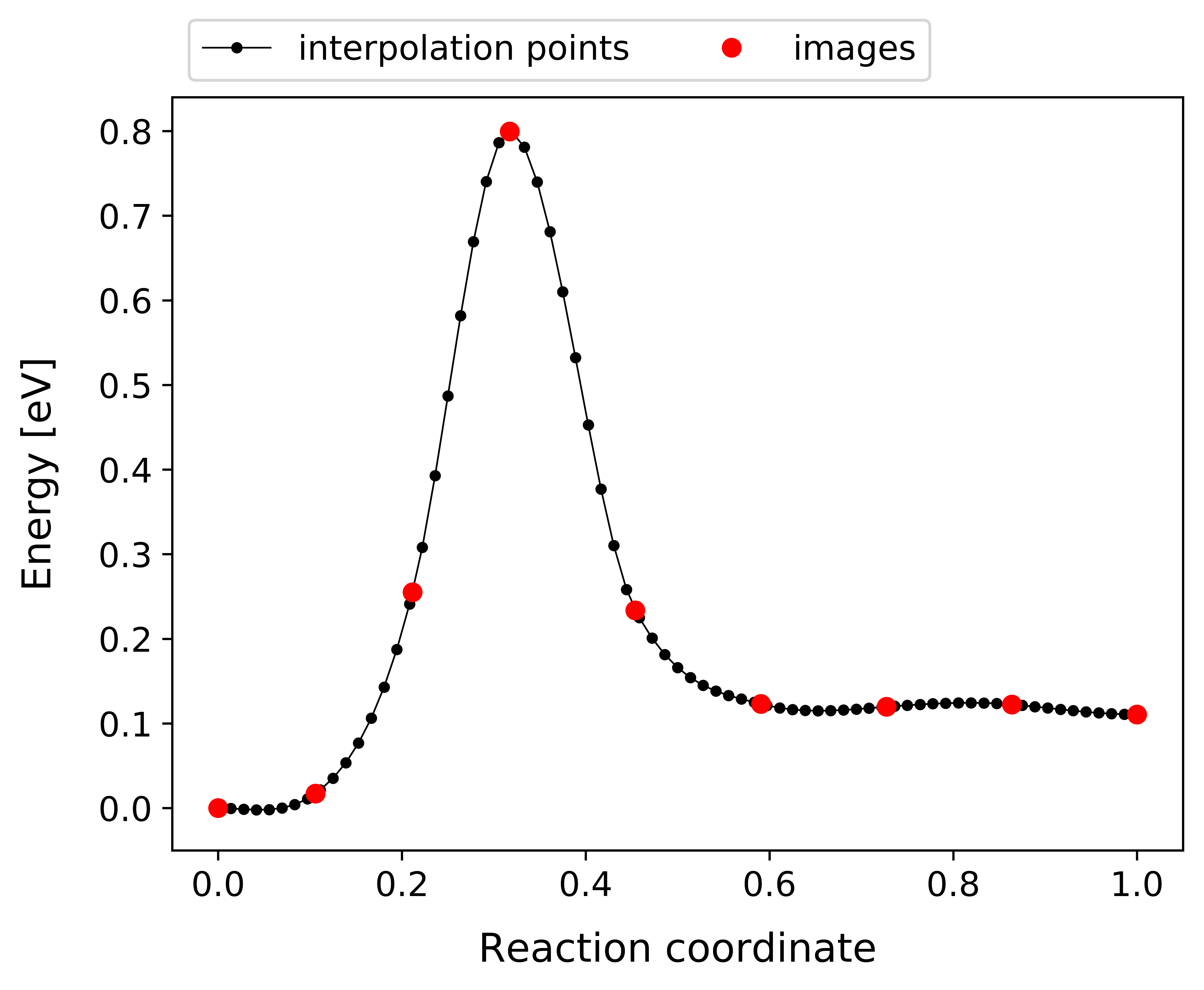} \hspace{0.2cm}
	\includegraphics[scale=0.39]{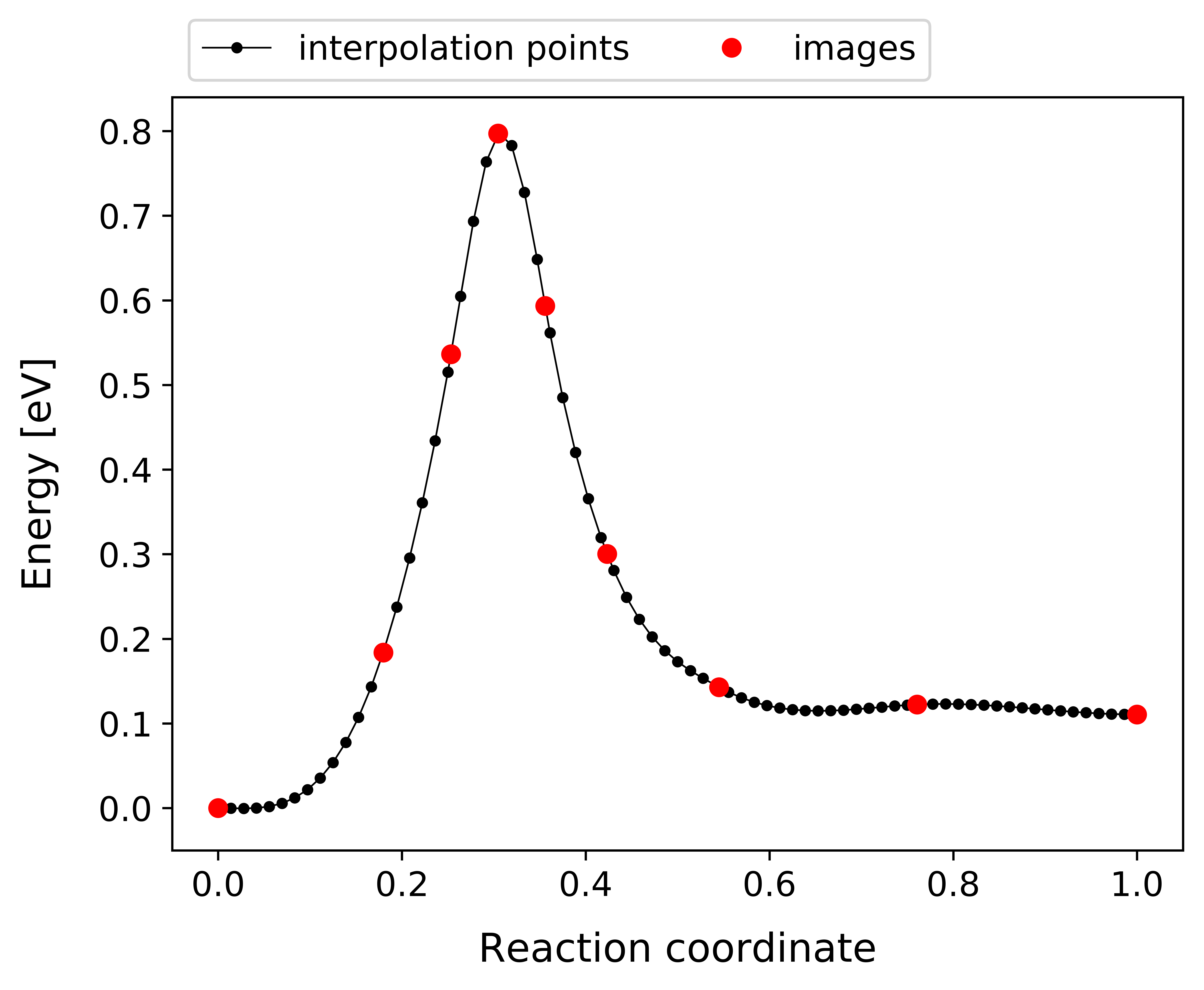}
	\caption{\small Minimum energy paths obtained with the basic {\footnotesize NEB} (left panel), the climbing image ({\footnotesize CI-NEB}, central panel) and the variable spring constants ({\footnotesize VARK-NEB}, right panel) methods, using the steepest descent (\textsc{Sd}) algorithm with an optimization step of $ds = 0.8$ Bohr$^2$/Ha to converge the reaction path towards the minimum energy path. Details about the activation barriers, as well as about the vibrational frequencies and the geometrical distances among the atoms involved in the proton exchange process for the transition state configurations are reported in Tables \ref{tab:chabazite_distances_ts} and \ref{tab:chabazite_de_nu}. The interpolation of the minimum energy paths is done using 72 points. The method used for the interpolation is described in Section \ref{sec:mep_interpolation}.}\label{fig:chabazite_reaction_paths}
\end{figure*}

The inability of the basic {\footnotesize NEB} method to capture the correct transition state is evident when looking at the profile of the {\small MEP} reported in the left panel of Figure \ref{fig:chabazite_reaction_paths}, as the number of images is not sufficient to correctly locate the transition state. On the other hand, both the {\small CI-NEB} and the {\small VARK-NEB} methods, characterized by an increase in the resolution around the saddle point (see central and right panels of Figure \ref{fig:chabazite_reaction_paths}, respectively) provide a good description of the transition state, even with $m = 9$ images along the whole path.

In order to better characterize the transition state in the case of the basic {\small NEB} method, a greater number of images has been used to discretize the reaction path. The results of these calculations are outlined in Table \ref{tab:chabazite_de_nu_more_images}, and the corresponding minimum energy paths obtained using the steepest descent path minimization method are reported in Figure 3
of the supplementary material, highlighting an increasing in the accuracy for the calculation of the activation energy as the number of the images increases.
However, even with a value of $m = 25$, the activation energy and especially the imaginary frequency associated to the transition state found by the basic {\small NEB} method are not in good agreement with the results obtained with the climbing image and the variable spring constants approaches, as well as with the theoretical literature values. In this case, the basic {\small NEB} method proves to be less accurate and less efficient in locating the transition state with respect to the other two improved methods, that are the climbing image and the variable spring constants strategies, with a much greater number of images required to correctly characterize the transition state.

\begin{table}[htb]
	\renewcommand{\arraystretch}{1.28}
	\setlength{\tabcolsep}{4pt}
	\begin{tabular}{ll|c|c|c}
		\multicolumn{5}{l}{Number of images $m = 9$ - {\footnotesize B3LYP} functional} \\
		\toprule
		\multicolumn{2}{c|}{Method} & $n_{opt}$ & $E_a$ [eV] & $\nu$ [cm$^{-1}$] \\
		\toprule
		\multirow{2}{*}{Basic {\footnotesize NEB}} & \textsc{Sd} & 1038 & 0.475143 (0.364577) & --- \\
		& \textsc{Cg} & 1035 & 0.475451 (0.364885) & --- \\
		\hline
		\multirow{2}{*}{{\footnotesize CI-NEB}} & \textsc{Sd} & 508 & 0.799548 (0.688982) & 1343.1208$\,i$ \\
		& \textsc{Cg} & 511 & 0.799556 (0.688990) & 1343.1214$\,i$ \\
		\hline
		\multirow{2}{*}{{\footnotesize VARK-NEB}} & \textsc{Sd} & 532 & 0.797144 (0.686577) & 1322.5652$\,i$ \\
		& \textsc{Cg} & 532 & 0.797199 (0.686633) & 1322.9498$\,i$ \\
		\toprule
		\multicolumn{3}{l|}{\footnotesize DRC} & 0.6504 [$a$] & 1218$\,i$ [$a$] \\
		\multicolumn{3}{l|}{\footnotesize DRC fixed cell} & 0.7935 [$a$] & 1334$\,i$ [$a$] \\
		\multicolumn{3}{l|}{\footnotesize DRC triple$\,$-$\,\zeta$} & 0.7415 [$a$] & 1308$\,i$ [$a$] \\
		\hline
		\multicolumn{3}{l|}{{\footnotesize QM-P}ot} & 0.7628 [$b$] & 1151$\,i$ [$b$] \\
		\toprule
	\end{tabular}
    \caption{\small Number of optimization steps $n_{opt}$ required to find the {\footnotesize MEP}, activation energy $E_a$ and imaginary frequency $\nu$ of the transition state. Different {\footnotesize NEB} methods (see Section \ref{sec:theory}) and various optimization schemes (see Section \ref{sec:path_minimization}) for the search of the minimum energy path are compared with other theoretical results. Results labeled with [$a$] are taken from Rimola \emph{et al.},\cite{chab_rimola_2010} with [$b$] from Sierka and coworkers.\cite{chab_sierka_2001} The optimization steps in the \textsc{Sd} and \textsc{Cg} path minimization algorithms are reported in the main text of Section \ref{sec:chabazite_results}. The basic {\footnotesize NEB} method is not able to locate a first-order saddle point, so that the corresponding frequency values for the transition state are left blank.}\label{tab:chabazite_de_nu}
\end{table}

\begin{table}[htb]
	\renewcommand{\arraystretch}{1.2}
	\setlength{\tabcolsep}{4pt}
	\begin{tabular}{lc|c|c|c}
		\multicolumn{5}{l}{Basic {\footnotesize NEB} - {\footnotesize B3LYP} functional} \\
		\toprule
		Images & \multicolumn{1}{c|}{Method} & $n_{opt}$ & $E_a$ [eV] & $\nu$ [cm$^{-1}$] \\
		\toprule
		\multirow{2}{*}{$m = 9$} & \textsc{Sd} & 1038 & 0.475143 (0.364577) & --- \\
		& \textsc{Cg} & 1035 & 0.475451 (0.364885) & --- \\
		\hline
		\multirow{2}{*}{$m = 12$} & \textsc{Sd} & 1760 & 0.697654 (0.587087) & 441.0805$\,i$ \\
		& \textsc{Cg} & 1766 & 0.697345 (0.586779) & 438.2428$\,i$ \\
		\hline
		\multirow{2}{*}{$m = 15$} & \textsc{Sd} & 565 & 0.649948 (0.539381) & 414.8259$\,i$ \\
		& \textsc{Cg} & 538 & 0.644330 (0.533763) & 388.3860$\,i$ \\
		\hline
		\multirow{2}{*}{$m = 18$} & \textsc{Sd} & 539 & 0.704545 (0.593979) & 744.4183$\,i$ \\
		& \textsc{Cg} & 544 & 0.705515 (0.594949) & 749.0149$\,i$ \\
		\hline
		\multirow{2}{*}{$m = 25$} & \textsc{Sd} & 537 & 0.749974 (0.639408) & 1050.4820$\,i$ \\
		& \textsc{Cg} & 541 & 0.748394 (0.637828) & 1040.1668$\,i$ \\
		\toprule
	\end{tabular}
	\caption{\small Number of optimization steps $n_{opt}$ required to find the {\footnotesize MEP}, activation energy $E_a$ and imaginary frequency $\nu$ of the transition state, obtained with the basic {\footnotesize NEB} method (see Section \ref{sec:theory}) using different number of images $m$ and various optimization schemes (see Section \ref{sec:path_minimization}) for the search of the minimum energy path. The optimization steps in the \textsc{Sd} and \textsc{Cg} path minimization algorithms are reported in the main text of Section \ref{sec:chabazite_results}.}\label{tab:chabazite_de_nu_more_images}
\end{table}

The excellent agreement between our results and the other theoretical calculations, at the {\small B3LYP} level, confirms the reliability of the improved {\small NEB} algorithms, implemented within the \textsc{Crystal} code, in computing the activation energy and finding the transition state for local transitions in periodic crystalline systems.


\section{Computational details}\label{sec:computational_details}

The new implementation of the {\small NEB} algorithm described in this work has been performed within a beta version of the \textsc{Crystal23} package for ab initio quantum chemistry and solid state physics.\cite{crystal23}
%
%

Geometry optimizations are performed using analytical gradients with respect to atomic coordinates, within a quasi-Newtonian algorithm combined with the Broyden-Fletcher-Goldfarb-Shanno ({\small BFGS}) formula.\cite{bfgs_1, bfgs_2, bfgs_3, bfgs_4, bfgs_5} 
The default convergence thresholds were adopted.\cite{crystal23,crystal23manual}


Harmonic frequencies, computed at the $\bm{\Gamma}$ point, are obtained through the diagonalization of the mass-weighted Hessian in Cartesian coordinates, where the second derivatives of the energy with respect to atomic displacements are evaluated numerically by means of a central-difference formula.\cite{hfreq_1, hfreq_2}
In the calculation of frequencies for the transition states, any residual symmetry has been removed, to ensure consistency with the {\small NEB} calculations performed within the P$_1$ space group. The self-consistent field energy threshold was set equal to $10^{-12}$ Ha per unit cell, for all the systems.

%
The {\small NEB} calculations for the reaction \eqref{eq:reaction1} were performed using an unrestricted Hartree-Fock Hamiltonian.
The Coulomb and exchange series, summed in direct space, are truncated using overlap criteria, with thresholds given by [$20, 20, 20, 20, 20$].\cite{crystal23manual}
The threshold for the self-consistent field algorithm convergence on the total energy is chosen equal to $10^{-8}$ Ha. 
An all electron basis set, consisting of contracted Gaussian-type atomic orbitals ({\small AOs}) functions, named 5-11 G$^*$, was used.\cite{H_5-11G*_dovesi_1984}
Details on the {\small NEB} parameters are reported in Section \ref{sec:collinear_proton_tranfer_results}.

The {\small NEB} calculations for the reaction \eqref{eq:reaction2} were performed in the framework of {\small DFT}, using the {\small B3LYP} exchange-correlation functional. The Coulomb and exchange series, summed in direct space, are truncated using overlap criteria with thresholds set to [$20, 20, 20, 20, 20$]. The threshold for the self-consistent field algorithm convergence on the total energy is fixed at $10^{-8}$ Ha.
The all-electron basis sets by C. Gatti $et \, al.$\cite{HCNO_gatti1994} were adopted for the different elements,
resulting in 57 {\small AOs} for the whole molecular system.
Details on the {\small NEB} parameters are reported in Section \ref{sec:formamide_results}.

The {\small NEB} calculations for the proton exchange process in the acidic chabazite were performed in the framework of {\small DFT}, using {\small B3LYP} exchange-correlation functional.
The {\small DFT} exchange-correlation contribution was evaluated by numerical integration over the unit cell volume, using a pruned grid,\cite{grid_1,grid_2} consisting of 75 radial points and a maximum number of 974 angular points in regions relevant for chemical bonding.\cite{crystal23manual} 
The diagonalization of the Fock matrix and the integration over the reciprocal space is carried out using the Monkhorst-Pack mesh,\cite{MP_mesh} consisting in a grid of 14 $\bm{k}$ points in the irreducible part of the first Brillouin zone.
The Coulomb and exchange series, summed in direct space, are truncated using overlap criteria, with thresholds given by [$6, 6, 6, 6, 14$].\cite{crystal23manual}
The threshold for the self-consistent field algorithm convergence on the total energy is chosen equal to $10^{-8}$ Ha. 
A set of polarized double$\,$-$\,\zeta$ quality type all-electron basis 
were used
for the different elements,
with a total of 557 {\small AOs} per unit cell.

Default values for the parameters controlling the {\small NEB} algorithm have been defined and used in all the tests.
In the basic and {\small CI-NEB} approaches, the value of the elastic constant was set to $k = 0.1$ Ha/Bohr$^2$.
The {\small CI-NEB} technique is activated after 10 {\small NEB} iterations, i.e. starting from the eleventh {\small NEB} step. 
The minimum and maximum values of the spring constants in the {\small VARK-NEB} approach are set equal to $k_{min} = 0.1$ Ha/Bohr$^2$ and $k_{max} = 0.5$ Ha/Bohr$^2$, respectively, and the method is active from the beginning of the simulation.
The values of the optimization steps adopted for each simulation is reported in Section \ref{sec:results}, for all the analyzed systems. 
The thresholds for the path minimization regarding ($i$) the Euclidean norm and ($ii$) the root-mean-square of the components of the real nuclear force vectors orthogonal to the path are 10$^{-3}$ Ha/Bohr and 5$\cdot 10^{-4}$ Ha/Bohr, respectively.

\section{Conclusions}\label{sec:conclusions}

This work documents the implementation of the {\small NEB} method within the \textsc{Crystal} code, a quantum mechanical ab initio program, capable of modeling systems with different periodicity, using localized Gaussian functions as basis set. This algorithm allows to compute reaction paths for molecular and periodic systems, using hybrid density functionals, opening the possibility to obtain accurate and efficient determination of minimum energy paths as well as activation barriers and transition state configurations in solid state reactions, surface processes, and chemical-physical transformations in molecular and extended periodic environments.
To validate the implementation, we analyzed the reliability and efficiency of different {\small NEB} variants
(namely, basic, {\small CI-NEB} and {\small VARK-NEB}) on representative molecular and bulk reactions.
The results demonstrate that:
($i$) the implementation is reliable, robust, versatile and general purpose;
($ii$) the parameter associated with the step for the path minimization has a strong impact on the convergence speed, while maintaining a good and consistent accuracy up to a reasonable high value;
($iii$) the {\small CI-NEB} and the {\small VARK-NEB} methods proved to be the most efficient in identifying the correct transition state, and
($iv$) the steepest descent algorithm for finding the {\small MEP} outperforms the conjugate gradient method.
The integration of the {\small NEB} procedure into the \textsc{Crystal} code, which is able to perform quantum mechanical hybrid density functional calculations on periodic systems at a reasonably low cost, lays the foundation for future investigations of complex chemical-physical processes in condensed phase systems, at a high level of theory.


\section{Supplementary Material}

The supplementary material provides additional theoretical details, implementation aspects of the computational method, a flowchart of the implemented algorithm, information about the optimization step setting used for path minimization of the studied reactions and processes, as well as data about the geometrical structures involved in the tautomerization of formamide, and figures of the minimum energy paths for the proton transfer process in chabazite bulk.

\section{Acknowledgments}\label{sec:acknowledgment}
This project has received funding within the European Union's Horizon {\small 2020} research and innovation program from the European Research Council ({\small ERC}) for the project “Quantum Chemistry on Interstellar Grains'' ({\small QUANTUMGRAIN}), grant agreement No. 865657.
Additional funding was provided by Project {\small CH4.0} under the {\small MUR} program “Dipartimenti di Eccellenza {\small 2023-2027}'' ({\small CUP: D13C22003520001}).
The Spanish {\small MICINN} is also acknowledged for funding the projects {\small PID2021-126427NB-I00} and {\small CNS2023-144902}. Albert Rimola acknowledges Accademia delle Scienze di Torino for supporting the project “In silico interstellar grain-surface chemistry'', and gratefully acknowledges support through {\small 2023 ICREA} Award.
The authors thank Francesca Capellino for granting permission to use an image from her bachelor thesis in the graphical abstract of this article.

\section{Data availability}
The data that support the findings of this study are available within the article and its supplementary material.


\section*{References}
\vspace{-0.4cm}
\bibliography{neb_article} 

\end{document}